\documentclass[a4paper,fleqn,usenatbib,useAMS]{mnras}
\pdfoutput=1

\usepackage{graphicx}	
\usepackage{amsmath}	
\usepackage{amssymb}	
\usepackage{multicol}        
\usepackage{bm}		
\usepackage{pdflscape}	

\newcommand{\kms}{\,km\,s$^{-1}$} 

\newcommand{\teff}{\ensuremath{T_{\rm eff}}}
\newcommand{\logg}{\ensuremath{\log g}}
\newcommand{\vsini}{\ensuremath{v\sin i}}
\newcommand{\cossam}{{\sc cossam}}
\newcommand{\cossamsimplex}{{\sc cossam\_simple}}
\newcommand{\sparti}{{\sc sparti}}
\newcommand{\spartisimplex}{{\sc sparti\_simple}}
\newcommand{\atlastwelve}{{\sc atlas12}}
\newcommand{\atlasnine}{{\sc atlas9}}
\newcommand{\Tlusty}{{\sc tlusty}}

\usepackage[T1]{fontenc}
\usepackage{ae,aecompl}

\usepackage{txfonts}

\title[Chemical Abundance Analysis of NGC\,6250]{A spectroscopic study of the open cluster NGC\,6250 \thanks{Based on observations made with European Southern Observatory Telescopes at the Paranal Observatory under programme ID 079.D-0178A.}}

\author[A.~J. Martin]{A.~J. Martin$^{1,2}$
\thanks{Contact e-mail: \href{mailto:ajm@arm.ac.uk}{ajm@arm.ac.uk}}\thanks{Present address: Armagh Observatory \& Planetarium, College Hill, Armagh, BT61~9DG, UK},
M.~J. Stift$^{1}$,
L. Fossati$^{3}$,
S. Bagnulo$^{1}$,
C. Scalia$^{4,5}$,
F. Leone$^{4,5}$,
\newauthor and B. Smalley$^{2}$\\
$^{1}$Armagh Observatory \& Planetarium, College Hill, Armagh, BT61 9DG, UK\\
$^{2}$Astrophysics Group, Keele University, Staffordshire ST5 5BG, UK\\
$^{3}$Space Research Institute, Austrian Academy of Sciences, Schmiedlstrasse 6, A-8042 Graz, Austria\\
$^{4}$Universit{\`a} di Catania, Dipartimento di Fisica e Astronomia, Sezione Astrofisica, Via S. Sofia 78, I-95123 Catania, Italy\\
$^{5}$INAF--Osservatorio Astrofisico di Catania, Via S. Sofia 78, I-95123 Catania, Italy}

\date{Accepted 2016 November 22. Received 2016 November 22; in original form 2016 June 10}

\pubyear{2016}

\begin{document}
\label{firstpage}
\pagerange{\pageref{firstpage}--\pageref{lastpage}}
\maketitle

\begin{abstract}
  We present the chemical abundance analysis of 19 upper
  main-sequence stars 
  of the young open cluster NGC\,6250
  ($\log\,t\,\sim\,7.42\,$yr).  This work is part of a project aimed at setting
  observational constraints on the theory of atomic diffusion in stellar photospheres, 
  by means of a systematic study
  of the abundances of the chemical elements of early F-, A- and late
  B-type stars of well-determined age. Our data set consists of low-,
  medium- and high-resolution spectra obtained with the Fibre Large Array Multi Element Spectrograph (FLAMES)
  instrument of the ESO Very Large Telescope (VLT). To perform our analysis, we have
  developed a new suite of software tools for the chemical abundance
  analysis of stellar photospheres in local thermodynamical 
  equilibrium. Together with the chemical
  composition of the stellar photospheres, we have provided new
  estimates of the cluster mean radial velocity, proper motion, 
  refined the cluster membership, and we have given the 
  stellar parameters including masses and fractional age. We 
  find no evidence of statistically significant correlation between 
  any of the parameters, including abundance and cluster age, except perhaps for 
  an increase in Ba abundance with cluster age. We have proven
  that our new software tool may be successfully used for the chemical
  abundance analysis of large data sets of stellar spectra.\\
\end{abstract}

\begin{keywords}
stars: abundances - open clusters and associations: individual: NGC 6250.
\end{keywords}


\section{Introduction}
The spectra of early F-, A- and late B-type stars frequently show a
wealth of signatures of various physical phenomena of comparable
magnitude, such as, for instance, pulsation, the presence of a
magnetic field and a non-homogeneous distribution of the chemical
elements \citep[e.g.,][]{Landstreet2004}. The latter is an effect of the diffusion of the chemical
elements, a mechanism that is particularly important to study because
 it affects the apparent chemical composition of stars. In principle,
effects of the diffusion that operate at a large time-scale (i.e.,
comparable to the stellar lifetime) may even mimic those due to the
Galactic chemical evolution.  Therefore, it is important to understand
whether the relative chemical composition of the photosphere appear
systematically different from that of younger stars.

In order to obtain information about time-dependent processes acting
in stellar photospheres, we have chosen to study stars that are member
of open clusters of various ages. This is because the age of an open
cluster may be determined with a much better accuracy than that of
individual stars in the field, in particular when the star is in the
first-half of its main-sequence lifetime
\citep[e.g.,][]{Bagnulo06}.  A second advantage is that open cluster
stars are presumably formed with the same chemical composition, so
that any difference in the observed chemical composition between
cluster members may be directly linked to one of the stellar properties
 (e.g., effective temperature and/or rotation).
 
 We have considered the low- and mid-resolution spectra of 32 stars observed with the FLAMES
instrument of the ESO VLT in the field of view of the open cluster NGC 6250, and 
we performed a detailed chemical abundance analysis of the 19 member stars. 
 Our observations
are part of a data set containing the spectra of approximately 
  1000 stars observed as part of a 
 larger effort 
 to explore how various physical
effects change as a function of stellar age, in particular to set observational 
constraints to
the theory of atomic diffusion in stellar photospheres \citep{Michaud1970}, both in the cases of magnetic
and non-magnetic atmospheres. The overall project includes data 
for potential members of various open clusters, covering ages from $\log t
= 6.8$ to 8.9 and distance moduli from 6.4 to 11.8. The full list of
observed open clusters is given by \citet{Fossati08a}. 

The analysis of three of these clusters has been performed 
by \citet{Kilicoglu2016} (NGC6405); \citet{Fossati07,Fossati08b,Fossati10} 
and \citet{Fossati11a} (Praesape cluster and NGC\,5460). \citet{Kilicoglu2016}
found NGC\,6405 to have an age of $\log t \sim 7.88$, 
a distance of 400\,pc $\pm$ 50\,pc and an [Fe/H] metallicity of  
0.07 $\pm$ 0.03. The Praesape cluster has an age of $\log t \sim 8.85 \pm 0.15$ \citep{Gonzalez_Garcia2006} and it is at a distance of 180\,pc $\pm$ 10\,pc \citep{Robichon1999}. 
\citet{Fossati11a} found NGC\,5460 to have an age of $\log t \sim 8.2 \pm 0.1$, 
a distance of  720\,pc $\pm$ 50\,pc and a near solar metallicity.

With an age of $\log t \sim 7.42$\,yr and a distance of 865\,pc
\citep{Kharchenko13}, NGC\,6250 is 
 both the youngest and most distant cluster analysed as part of 
 this project so far. 
  Further studies completed by different groups, with data that can be used as 
part of this study, include those by 
\citet{Gebran08a} and  
\citet{Gebran08b,Gebran10} (Coma Berenices, $\log t$ = 8.65; 
the Pleiades, $\log t$ = 8.13; 
and Hyades, $\log t$ = 8.9); \citet{Folsom2007} 
and \citet{Villanova09} (NGC\,6475, $\log t$ = 8.48); and 
 \citet{Stutz2006} (IC 2391, $\log t$ = 7.66).
 
 The study by \citet{Bailey2014} searched for trends between 
 chemical abundance and stellar parameters of chemically 
 peculiar Ap stars, to determine 
 whether chemical peculiarities change as a star evolves. This 
 data will allow us to compare the behaviour of 
 chemically peculiar magnetic 
 stars with our sample of chemically normal stars.

To analyse the remaining clusters for 
this project in a more efficient manner, and in particular
to deal with the especially interesting case of magnetic stars, we have
developed \sparti\ (SpectroPolarimetric Analysis by Radiative Transfer
Inversion), a software tool based on the radiative transfer code
\cossam\ \citep{Stift00,Stift12}. \sparti\ will be presented in a
forthcoming paper (Martin et al., in preparation). In this work, we introduce
its simplest version, \spartisimplex, specifically designed to deal
with the non-magnetic case. \spartisimplex\ is based around
\cossamsimplex, which in turn is a modified version of the code
\cossam\ for the spectral synthesis of magnetic atmospheres.
This approach has the advantage that both magnetic and non-magnetic
stars may be analysed in a homogeneous way. Eventually, the comparison
of the chemical composition of magnetic and non-magnetic stars
belonging to the same cluster will allow us a more accurate analysis
of the effects of magnetic fields on the diffusion of the chemical
elements in a stellar photosphere. Our new software suite is fully
parallelized, which reduces the CPU time required to analyse each star.

In this paper, we first describe \cossamsimplex\ and \spartisimplex\
(Sections~\ref{Sect_Cossam} and \ref{Sect_Spartisimple}), then we present
the observations (Section~\ref{Sect_Observations}), we establish cluster
membership (Section~\ref{Sect_Membership}) and we determine the fundamental parameters of the cluster
members (Section~\ref{Sect_Fundamental}). We then present new spectroscopic
observations of the cluster NGC\,6250 (Section~\ref{Sect_Observations}). 
Finally we present and discuss our results (Section~\ref{Sect_Results}). 
Our conclusions are summarized in
Section~\ref{Sect_Conclusions}.
\section{The Cossam code}\label{Sect_Cossam}

\cossam, the `Codice per la sintesi spettrale nelle atmosfere
magnetiche' is an object-oriented and fully parallelized polarized
spectral line synthesis code, under GNU copyleft since the year 2000.
It allows the calculation of detailed Stokes {\it IQUV} spectra in the
Sun and in rotating and/or pulsating stars with dipolar and quadrupolar
magnetic geometries. Software archaeology reveals that {\sc cossam} harks
back to the {\sc algol\,60} code {\sc  analyse\,65} by \citet{Baschek66} and
to the {\sc fortran} code {\sc adrs3} by \citet{Chmielewski79}. {\sc cossam} is the
first code of its kind that takes advantage of the sophisticated
concurrent constructs of the {\sc ada} programming language that make it
singularly easy to parallelize the line synthesis algorithms without
having recourse to message passing interfaces. `Tasks', each of which
has its own thread of control and each of which performs a sequence
of actions -- such as opacity sampling and solving the polarized
radiative transfer equation over a given spectral interval -- can
execute concurrently within the same program on a large number of
processor cores. Protected objects, which do not have a thread of
control of their own, are accessed in mutual exclusion (i.e. only
one process can update a variable at a time) and provide efficient
synchronisation with very little overhead.
\subsection{Physics and numerics}
\cossam\ assumes a plane-parallel atmosphere and local
thermodynamic equilibrium (LTE). It is convenient to use the VALD
data base \citep{Piskunov95} extracting atomic transition data
including radiation damping, Stark broadening and van der Waals
broadening constants.  The atomic partition functions are calculated
with the help of the appropriate routines in \atlastwelve\
\citep{Kurucz05}. Land{\'e} factors and $J$-values for the lower and the
upper energy levels provided by VALD make it possible to determine the
Zeeman splitting and the individual component strengths of each line;
in the case the Land{\'e} factors are missing, a classical Zeeman triplet
is assumed. For the continuous opacity $\kappa_{\rm c}$ at a given
wavelength, \cossam\ employs \atlastwelve\ routines \citep{Kurucz05}
rewritten in {\sc ada} by \citet{Bischof05}. The total line opacities
required in the formal solver are determined by full opacity sampling
of the $\sigma_{-}$, $\sigma_{+}$ and ${\bm \pi}$ components
separately. The opacity profiles -- Voigt and Faraday functions -- of
metallic lines are based on the rational expression found in
\citet{Hui78}.  The approximation to the hydrogen line opacity
profiles given in \Tlusty\ \citep{Hubeny95} has proved highly
satisfactory and easy to implement. The higher Balmer series members
are treated according to \citet{Hubeny94}; this recipe is based on the
occupation probability formalism \citep{Dappen87,Hummer88,Seaton90}.

By default, \cossam\ employs the Zeeman Feautrier method \citep{Auer77}, 
reformulated by \citet{Alecian04} in order to treat blends
in a static atmosphere. Alternatively, the user can choose the somewhat
faster but less accurate DELO method \citep{Rees89}. Since most of
the CPU time is spent on opacity sampling, the overall cost of Zeeman
Feautrier is only slightly higher compared to DELO. In the local (`solar')
case, the emerging Stokes spectrum is calculated for one given point on
the solar surface  -- specified by the position $\mu = \cos \theta =
(1 - r^2)^{1/2} $ -- and the attached magnetic vector. In the `stellar'
(disc-integrated) case, \cossam\ has to integrate the emerging
spectrum over the whole visible hemisphere, i.e. over $0 < \mu \leq 1$,
taking into account rotation, (non-)radial pulsation and a global dipolar
or quadrupolar magnetic field structure. 

Different spatial grids are
provided to best cater for the different spectral line synthesis
problems usually encountered. We may distinguish {\em corotating} from
{\em observer-centred} grids. The former are extensively used in Doppler
mapping (see e.g. \citet{Vogt87}); the entire stellar surface is
split into elements of approximately equal size. Chemical and/or magnetic
spots can easily be modelled with the help of these corotating spatial
grids. Observer-centred grids usually are of a fixed type where neither
the magnetic field geometry nor rotation and/or pulsation determine the
distribution of the quadrature points. \cossam\ also provides a third
type of grid, namely an adaptive grid as discussed in \citet{Stift85} and
\citet{Fensl95}. A special algorithm provides optimum 2D-integration by
ensuring that the change in the monochromatic opacity matrix between two
adjacent quadrature points does not exceed a certain percentage. The
point distribution can become very non-uniform, depending on the
direction of the magnetic field vector, its azimuth, the Doppler shifts
due to rotation and/or pulsation, and on the amount of limb darkening.
At the same level of accuracy of the resulting Stokes profiles, it is
thus possible to greatly reduce the number of quadrature points compared
to fixed grids.
\subsection{{\sc cossam\_simple}, a fast tool for
chemical abundance analysis in non-magnetic stars}\label{Sect_Cossamsimple}
In principle, the original version of \cossam\ may naturally deal with
the non-magnetic case, just by setting the magnetic field strength to
zero. Practically, \cossam\ would still perform a number of
time-consuming numerical computations ending into flat-zero Stokes
$QUV$ profiles.
Therefore, the original code was modified to
take advantage of the various symmetries and simplifications of the 
non-magnetic case, and to allow extremely fast, but nevertheless
highly accurate integration of intensity profiles even in rapidly
rotating stars. \cossamsimplex\ calculates local spectra at high
wavelength resolution at various positions $\mu = \cos \theta$; the
stellar spectrum is then derived by integration over the appropriately
shifted local spectra. Instead of the hundreds or even thousands of
local spectra to be calculated for the general-purpose 2D-grid, a few
dozen local spectra prove sufficient.

After calculating a synthetic spectrum, we convolve it with a 
Gaussian matching the instrument resolution, the wavelength sampling 
of the resulting spectrum is then matched to the observed wavelength 
grid to allow for comparison.
\section{Inversion Method}\label{Sect_Spartisimple}
\spartisimplex\ is an inversion code that uses \cossamsimplex\ to
calculate synthetic stellar spectra, and the Levenberg--Marquardt
algorithm (LMA) to find the best-fitting parameters to observed stellar
spectra. Its free parameters are the chemical abundances of an
arbitrary number of chemical elements (including independent
abundances for each ionization stage), assuming a fixed model for the
stellar atmosphere (hence fixed values of effective stellar
temperature \teff\ and gravity \logg).
\subsection{Levenberg--Marquardt algorithm}
\label{sec:LMA}
The Levenberg--Marquardt algorithm  
\citep[LMA;][]{Levenberg44,Marquardt63}
 is a least-squares technique
that combines the Gauss--Newton and gradient-descent methods. It
allows one to determine the minimum of a multivariate function minimizing
the expression
\begin{equation} 
\chi^2 = \sum\limits_{i=1}^{n} \frac{\left[F_{\rm mod}
\left(\lambda_i,{\bm x}\right)-F_{\rm obs}\left(\lambda_i\right)\right]^2}{\sigma_i^2},
\label{eq:chi2} 
\end{equation} 
where $F_{\rm obs}$ is the observed spectrum, $\sigma$ is the error
associated with each spectral bin $i$, $F_{\rm mod}$ is the synthetic
spectrum, convolved with the instrument response, ${\bm x}$ is the array of free parameters
assumed in the spectral synthesis,
and $n$ is the number of spectral points.

\spartisimplex\ initially calculates a synthetic spectrum
with the chemical abundances set to solar values from
\citet{Asplund09}.  Initial estimate values of the projected equatorial
velocity $v\sin i$, the radial velocity $v_{\rm rad}$ and the
microturbulence $v_{\rm mic}$ are also required. After the initial
spectrum is generated, the LMA
 is run and convergence to the best solution is
usually reached within four to eight iterations. The algorithm is stopped
when the following conditions are met \citep{Aster2013}
\begin{equation} 
\begin{aligned}
\sum\limits_{i=1}^n [\nabla \chi^2]^2 &< \sqrt{\epsilon}\sum\limits_{i=1}^n [1+f(i)]\\
\sum\limits_{i=1}^n \left[x(i) - x_0(i)\right]^2 &< 
\sqrt{\epsilon}\sum\limits_{i=1}^n \left[1 + x(i)^2\right]\\
\sum\limits_{i=1}^n \left[F(i) - F_0(i)\right]^2 &< 
\epsilon\sum\limits_{i=1}^n \left[1 + F(i)^2\right],
\end{aligned}
\end{equation}
where $x_0$ is the previous parameter set, $F_0$ is the model spectrum
calculated with $x_0$ and $\epsilon = 10^{-4}$. The maximum number of iterations is
set to 50, if this number is reached, we reassess the starting 
parameters and rerun \spartisimplex.
\subsubsection{Semi-automatic identification of the chemical elements}
To quickly check which elements may be identified in the
observed spectrum, after the best-fit is found, we re-calculate
a number of synthetic spectra, each of which obtained after
setting to zero the abundance of a single element. We compare each of these new
synthetic spectra with the observed one and we check if the
reduced $\chi^2$ has varied by more than the signal-to-noise (S/N) threshold. 
If there is no change we consider that the 
element cannot be measured in the observed spectrum. If $\chi^2$ 
has varied, we check the spectrum to determine whether the element has 
visible spectral lines.
\subsection{Abundance uncertainties}
\label{sec:Uncer}
We estimate the uncertainties of the best-fitting parameters, including $v_{\rm mic}$ and 
$v\sin i$ by taking
the square root of the diagonal values of the covariance matrix
\begin{equation} 
{\rm cov}({\bm{\widehat{x}}}) = s^2{\bm{\Delta F \Delta F}}^T,
\label{eq:cov} 
\end{equation} 
where $\bm{\widehat{x}}$ is the vector of best-fit  and  
\begin{equation} 
s = \sqrt{\frac{\sum\limits_{i=0}^n\left[F_{\rm mod}\left(\lambda_i\right)-F_{\rm obs}\left(\lambda_i\right)\right]^2}{m-n}},
\label{eq:s} 
\end{equation} 
where $n$ is the number of wavelength points in the spectrum and $m$ the 
number of best-fitting parameters. 

The error so estimated is only a lower limit since
\spartisimplex\ assumes fixed values of $T_{\rm eff}$ and $\log
g$. Therefore, the covariance matrix does not contain information on
the effects of the uncertainties of $T_{\rm eff}$ or $\log g$. To take
these uncertainties into account, we run the inversion another four times,
setting $\teff = T^0_{\rm eff} \pm \Delta T_{\rm
  eff}$ and $\logg  = \log g^0\pm\Delta \log g$, where $\teff^0$ and
$\logg^0$ are our best estimates for \teff\ and \logg, respectively,
and $\Delta \teff$ and $\Delta \logg$ are their errors. As abundance
uncertainty we finally adopt
\begin{equation} 
\sigma = \sqrt{\sigma_{\rm cov}^2 + \sigma_{T_{\rm eff}}^2 +\sigma_{\log g}^2}.
\label{eq:err} 
\end{equation} 
where $\sigma_{\rm cov}$ is the error values calculated from equation~(\ref{eq:s}),
$\sigma_{T_{\rm eff}}$ is half of the difference between the best-fitting values for
the abundances obtained assuming $\teff = \teff^0 + \Delta \teff$  and
$\teff = \teff^0 - \Delta \teff$, and $\sigma_{\log g}$  half of the
difference between the best-fitting values for the abundances obtained
assuming $\logg = \logg^0 + \Delta \logg$  and
$\logg = \logg^0 - \Delta \logg$.
\subsection{Test cases: the Sun, HD\,32115 and 21\,Peg}
In order to check the full consistency between the results obtained
with \spartisimplex\ and those obtained in the previous spectral
analysis of open cluster members, we
have performed a test spectral analysis of the Sun, HD\,32115 and the star 21\,Peg.
We chose these stars because they represent three different regimes: 
in the solar spectrum, the Balmer lines are only sensitive to temperature; 
in the spectrum of HD\,32115, the Balmer lines are sensitive to temperature and 
surface gravity; and in the spectrum of 21 Peg the Balmer lines are more 
sensitive to surface gravity. 
 We 
 calculate the fundamental parameters for each star 
shown in Tables 
\ref{table:Solar}--\ref{table:21peg}. 
Good agreement is seen between the parameters we 
calculated and the previously published results. The higher 
value of $v \sin i$ that we measure for the Sun is due to the 
fact that we do not account for macroturbulence broadening, while 
the technique used to calculate the value of $v \sin i$ given in 
\citet{Prsa2016} does. For HD\,32115, 
we derived the \teff\ and \logg\ from the Balmer lines, 
since \citet{Fossati2011b}  showed the ionization balance of Fe is 
not sufficient in this star to determine $\log g$ accurately.

The $v_{\rm mac}$ of the Sun ranges from 1 to 4\,kms$^{-1}$; 
however, we convolved the solar spectrum to a resolution 
of 25900 to simulate the analysis of our GIRAFFE spectra. At 
this resolution, the effect of $v_{\rm mac}$ is not visible in our spectra. 
The analysis of the chemical abundances of the Sun shows 
good agreement with the results of \citet{Asplund09} within the 
error bars shown in Table \ref{table:Solar2}. 

To perform the comparison between the chemical abundances 
determined by \citet{Fossati09}  and those determined by 
{\sc sparti\_simple} for 21 Peg, we choose the same lines, atomic parameters 
and \atlastwelve\ model atmosphere as \citet{Fossati09} and removed
any lines that show NLTE effects and/or have evidence of a
hyperfine structure.

\begin{figure}
\centering
   \includegraphics[width=\columnwidth]{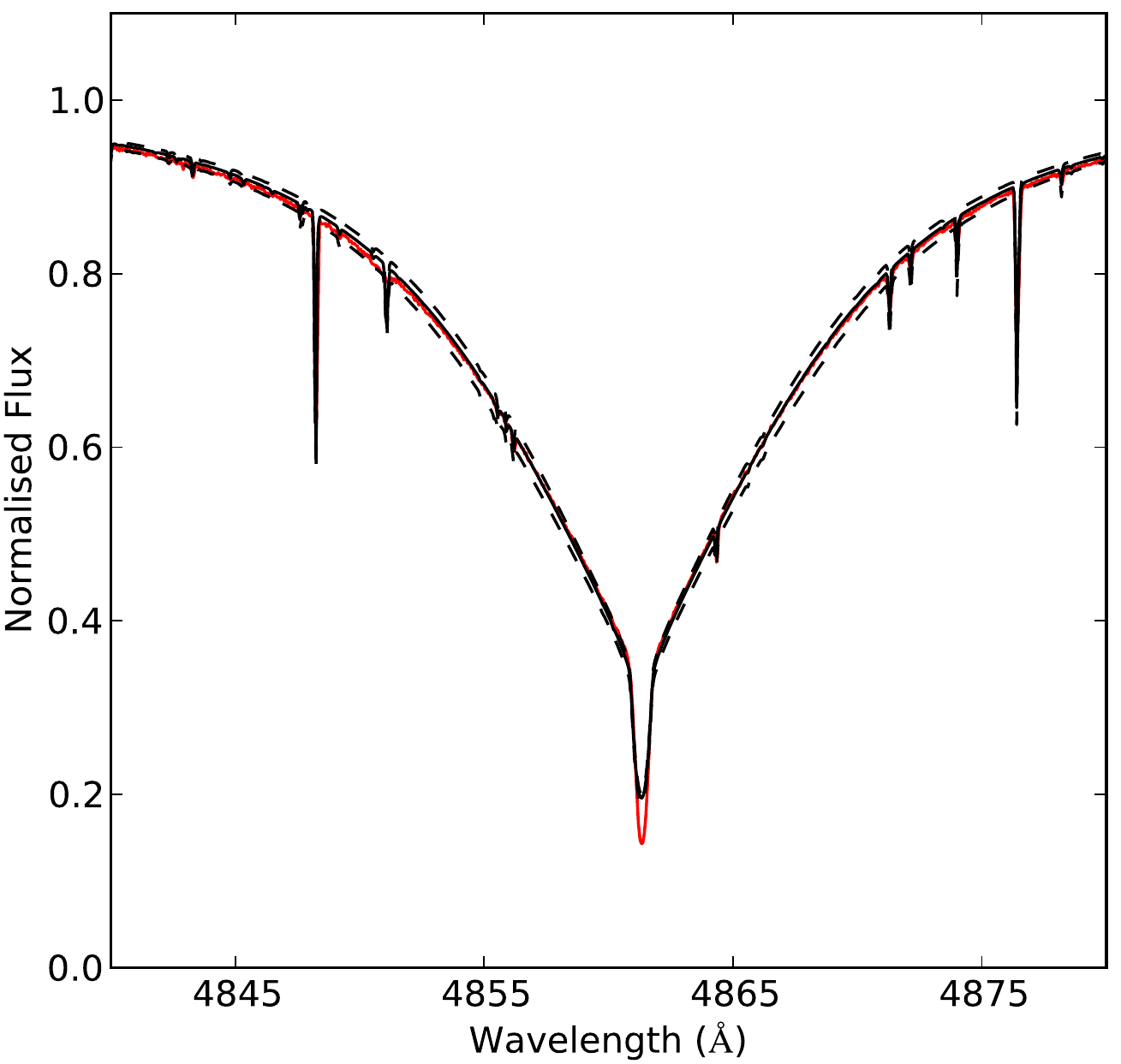}
     \caption{The fit of the H\,$\beta$ line in the spectrum of 21\,Peg. The solid red line is the observed spectrum. The dashed black is the synthetic spectrum calculated using $T_{\rm eff}=10400$\,K and $\log g = 3.55$. The two dashed blue lines are $\pm$ 200K}
     \label{fig:21Peg}
\end{figure}

The results of our test are shown in Fig.~\ref{fig:21Peg} (which shows
a comparison between our newly calculated H\,$\beta$ profile and
the observed spectrum) and in Table~\ref{table:21peg2}.  The numerical
results of Table~\ref{table:21peg2} demonstrate that our results agree
within the errors of those calculated by \citet{Fossati09}. 
\begin{table}
	\caption{The fundamental parameters associated with the Sun. Calculated 
	both by \citet{Prsa2016} and by the method used 
	in this work.}   
	\label{table:Solar}     
	\centering
	\begin{tabular}{l r@{ $\pm$ }l r@{ $\pm$ }l}
		\hline\hline\\[-2.0ex]   
		Fundamental parameters 	& \multicolumn{2}{c}{\citet{Prsa2016}}	& \multicolumn{2}{c}{This work}\\ 
		\hline\\[-2.0ex]
		$T_{\rm eff}$ (K)			& \multicolumn{2}{c}{5777}	& 5800	& 200\\
		$\log g$ (cgs)				& \multicolumn{2}{c}{4.438}	& 4.49	& 0.1\\
		$v\sin i$ (km s$^{-1}$)		& \multicolumn{2}{c}{1.2}	& 2.6 	& 0.1		\\
		$v_{\rm mic}$ (km s$^{-1}$)	& \multicolumn{2}{c}{0.875}	& 1.0 	& 0.03		\\
		\hline\\[-2.0ex]  
	\end{tabular}
\end{table}
\begin{table}
	\caption{The fundamental parameters associated with HD32115. Calculated 
	both by \citet{Fossati2011b} and by the method used 
	in this work.}   
	\label{table:HD32115}     
	\centering
	\begin{tabular}{l r@{ $\pm$ }l r@{ $\pm$ }l}
		\hline\hline\\[-2.0ex]   
		Fundamental parameters 	& \multicolumn{2}{c}{\citet{Fossati2011b}}	& \multicolumn{2}{c}{This work}\\ 
		\hline\\[-2.0ex]
		$T_{\rm eff}$ (K)			& 7250	& 100	&  7300	& 200\\
		$\log g$ (cgs)				& 4.2	& 0.1	&  4.2	& 0.1\\
		$v\sin i$ (km s$^{-1}$)		& 8.3	& 0.5	& 14.21 	& 0.07	\\
		$v_{\rm mic}$ (km s$^{-1}$)	& 2.5	& 0.2	& 2.5 	& 0.02		\\
		\hline\\[-2.0ex]  
	\end{tabular}
\end{table}
\begin{table}
	\caption{The fundamental parameters associated with the 21 Peg. Calculated 
	both by \citet{Fossati09} and by the method used 
	in this work.}   
	\label{table:21peg}     
	\centering
	\begin{tabular}{l r@{ $\pm$ }l r@{ $\pm$ }l}
		\hline\hline\\[-2.0ex]   
		Fundamental Parameters 	& \multicolumn{2}{c}{\citet{Fossati09}}	& \multicolumn{2}{c}{This work}\\ 
		\hline\\[-2.0ex]
		$T_{\rm eff}$ (K)		& 10400	& 200	& 10400	& 200\\
		$\log g$ (cgs)			& 3.55	& 0.1	& 3.5	& 0.1\\
		$v\sin i$ (km s$^{-1}$)	& 3.76	& 0.35	& 4.2	& 0.1		\\
		$v_{\rm rad}$ (km s$^{-1}$)	& 0.5	& 0.5	& 0.40	& 0.04	\\
		$v_{\rm mic}$ (km s$^{-1}$)	& 0.5	& 0.5	& 0.4	& 0.2		\\
		\hline\\[-2.0ex]  
	\end{tabular}
\end{table}
\begin{table}
	\caption{A comparison between the chemical abundances of the Sun calculated by \citet{Asplund09} and those calculated by {\sc sparti\_simple}.}  
	\label{table:Solar2}     
	\centering
	\begin{tabular}{l r@{ $\pm$ }l r@{ $\pm$ }l r}
		\hline\hline\\[-2.0ex]   
					& \multicolumn{5}{c}{$\log\left(N/{\rm H}\right)$}\\
		 			& \multicolumn{2}{c}{Asplund et al.}	& \multicolumn{2}{c}{{\sc Sparti}}  \\
		Element		& \multicolumn{2}{c}{(2009)}		& \multicolumn{2}{c}{{\sc Simple}} & $\Delta$\\
		\hline\\[-2.0ex]  
			Mg	& 7.60	& 0.04	& 7.63	& 0.02	& -0.03\\
			Ti  	& 4.95	& 0.05	& 4.96	& 0.01	& -0.01\\
			V  	& 3.93	& 0.08	& 3.97	& 0.01	& -0.04\\
			Cr 	& 5.64	& 0.04	& 5.60 	& 0.01	&  0.04\\ 
			Fe 	& 7.50	& 0.04	& 7.49 	& 0.01	&  0.01\\
			Ni 	& 6.22	& 0.04	& 6.21 	& 0.01	&  0.01\\
		\hline\\[-2.0ex]  
	\end{tabular}
\end{table}
\begin{table}
	\caption{A comparison between the chemical abundances of 21\,Peg calculated by \citet{Fossati09} and those calculated by {\sc sparti\_simple}, along with the solar abundances from \citet{Asplund09}.}   
	\label{table:21peg2}     
	\centering
	\begin{tabular}{l r@{ $\pm$ }l r@{ $\pm$ }l r r}
		\hline\hline\\[-2.0ex]   
					& \multicolumn{5}{c}{$\log\left(N/{\rm H}\right)$}\\
		 			& \multicolumn{2}{c}{Fossati et al.}	& \multicolumn{2}{c}{{\sc sparti}}  \\
		Element		& \multicolumn{2}{c}{(2009)}		& \multicolumn{2}{c}{{\sc simple}} & $\Delta$ & Solar\\
		\hline\\[-2.0ex]  
O {\sc i}	& 8.76	& 0.11	& 8.80	& 0.06	& -0.04	& 8.69\\
Al {\sc ii}	& 6.34	& 0.10	& 6.37	& 0.16	& -0.03	& 6.45\\
Si {\sc ii}	& 7.55	& 0.13	& 7.51	& 0.07	& 0.04	& 7.51\\
S {\sc ii}	& 7.18	& 0.13	& 7.24	& 0.34	& -0.06	& 7.12\\
Sc {\sc ii}	& 2.67	& 0.10	& 2.59	& 0.23	& 0.08	& 3.15\\
Ti {\sc ii}	& 4.81	& 0.09	& 4.78	& 0.03	& 0.03	& 4.95\\
V {\sc ii}	& 4.06	& 0.06	& 3.96	& 0.05	& 0.10	& 3.93\\
Cr {\sc ii}	& 5.84	& 0.10	& 5.79	& 0.02	& 0.05	& 5.64\\
Fe {\sc ii}	& 7.54	& 0.12	& 7.52	& 0.01	& 0.02	& 7.50\\
Ni {\sc ii}	& 6.43	& 0.09	& 6.38	& 0.03	& 0.05	& 6.22\\
Sr {\sc ii}	& 2.94	& -	& 2.91	& 0.04	& 0.03	& 2.87\\
Ba {\sc ii}	& 2.85	& 0.06	& 2.81	& 0.24	& 0.04	& 2.18\\
		\hline\\[-2.0ex]  
	\end{tabular}
\end{table}
\begin{table}
	\caption{Instrument settings information with the useful spectral lines
	              given for stellar \teff\ between $\sim 6000$\,\AA\
	              and $\sim 25000\,$\AA}   
	\label{table:Settings}     
	\centering
	\begin{tabular}{l l c l}
		\hline\hline\\[-2.0ex]
		Instrument	& Resolving	& Spectral				& Important\\
		setting		& power		& region\,($\AA$)		& spectral lines \\
		\hline\\[-2.0ex]
		LR3				& 7500			& 4500--5077	& H$\,\beta$\\
		HR9B 			& 25900			& 5139--5355	&Fe-peak (inc. Fe,Ti and Cr) \\
		&&&Mg triplet at $\lambda$'s \\
		&&&$\sim$ 5167, 5172 and 5183\,$\AA$\\
		HR11			& 24200			& 5592--5838	& Fe, Na and Sc \\
		LR6				& 8600			& 6438--6822	& H\,$\alpha$\\
		UVES				& 47000			& 4140--6210	& H\,$\beta$, H\,$\gamma$\\
						&				&			& Fe-peak (inc. Fe,Ti and Cr) \\
		&&&Mg triplet at $\lambda$'s \\
		&&&$\sim$ 5167, 5172 and 5183\,\AA\\
		\hline\\[-2.0ex]
	\end{tabular}
\end{table}
\begin{table*}
	\caption{List of programme stars. The proper motion 
	($\mu$) in right ascension (RA) and declination (DEC), for each star, is taken from 
	the UCAC2 catalogue \citep{Zacharias04}. The radial velocities ($v_{\rm r}$) are calculated using the method 
	in Section \ref{sec:RadV} and for the member stars using \spartisimplex. 
	$\Delta$ is the number of standard deviations the proper motions and radial 
	velocity are away from the cluster mean. 
	The S/N (column 8) corresponds to the 
	S/N per spectral bin of either the GIRAFFE settings or the UVES 520\,nm 
	setting. The $BV$ photometry is taken from the APASS catalogue \citep{Henden2016}  and the 
	$JK$ photometry is taken from the UCAC2 catalogue \citep{Zacharias04}. The Memb. column gives the results of our membership analysis, where `y' means the star is a member, `n' means the star is not a member. The final column gives the membership probabilities given by \citet{Dias06}.}
	\label{Tab:Prog}     
	\centering
	\resizebox{\textwidth}{!}{\begin{tabular}{lr@{$\pm$}l c r@{$\pm$}l c r@{$\pm$}lccccccccc}
		\hline\hline
		Star &
		\multicolumn{2}{c}{$\mu_{\rm RA}$}    & 
		$\Delta_{\mu_{\rm RA}}$ &
		\multicolumn{2}{c}{$\mu_{\rm DEC}$} &
	 	$\Delta_{\mu_{\rm DEC}}$ &
		\multicolumn{2}{c}{$v_{\rm r}$} &
		$\Delta_{v_{\rm r}}$ & S/N & \multicolumn{4}{c}{Phot. (Mag)}&& Dias\\
		name & 
		\multicolumn{2}{c}{$(\rm masyr^{-1})$}    && 
		\multicolumn{2}{c}{$(\rm masyr^{-1})$}    &&
		\multicolumn{2}{c}{$(\rm km s^{-1})$}      && LR3/HR9B/HR11/LR6 &$B$&$V$&$J$&$K$& Memb. & Memb.\\
		\hline
		CD-4511088 	& 2.4 &1.4 &0.6 &$-$3.5 &1.4 &0.4 &$-$36.0 & 0.5 & 2.6 &94/95/128/122 & 11.4 &11.0 &10.4 &10.1 &n & \\
		HD\,152706 	& $-$2.9 &1.4 &1.0 &$-$3.3 &1.3 &0.5 &$-$4.6 & 0.4 & 0.5 & 68[UVES] & 10.4 &10.1 &9.7 &9.5 &y & \\
		HD\,152743 	& 0.3 &1.4 &0.0 &$-$4.6 &1.4 &0.1 &0.4 & 52.4 & 1.0 &116/252/146/194 & 9.2 &9.1 &8.7 &8.6 &y & \\
		HD\,329261 	& $-$10.0 &0.5 &3.3 &$-$19.0 &0.5 &4.7 &$-$3.1 & 0.5 & 0.7 &59/62/84/128 & 11.2 &10.8 &9.9 &9.6 &n & \\
		HD\,329268 	& $-$6.0 &0.5 &2.0 &$-$7.0 &0.5 &0.7 &$-$20 & 3.0 & 1.0 &62/69/92/72 & 12.0 &11.4 &10.6 &10.3 &n & \\
		HD\,329269 	& 4.0 &0.5 &1.1 &$-$13.0 &0.5 &2.7 &$-$16 & 3.0 & 0.6 &58/60/84/118 & 11.7 &11.2 &9.9 &9.6 &n & \\
		NGC\,6250-11 	& 17.5 &1.4 &5.3 &$-$23.6 &1.4 &6.3 & 1 & 3.0 & 1.1 &51/62/75/72 & 13.4 &12.7 &11.4 &11.0 &n & \\
		NGC\,6250-13 		& 2.7 &2.5 &0.7 &$-$3.1 &2.4 &0.6 &$-$14 & 3.0 & 0.4 &49/59/70/58 & 13.5 &13.1 &12.0 &11.6 &y & \\
		TYC\,8327-565-1 & $-$0.5 &1.7 &0.3 &$-$1.6 &1.7 &1.1 &$-$9.4 & 0.2 & 0.1 & 65[UVES]& 11.0 &10.8 &10.5 &10.4 &y & \\
		UCAC\,12065030 & $-$2.4 &5.2 &0.9 &$-$17.9 &5.2 &4.4 &$-$20 & 3.0 & 1.0 &43/53/62/49 & 14.2 &13.5 &11.8 &11.4 &n & 47\\
		UCAC\,12065057 & $-$14.4 &2.6 &4.6 &$-$21.2 &2.4 &5.5 &$-$49 & 3.0 & 3.9 &50/59/72/61 & 13.6 &12.9 &11.4 &10.9 &n & 100\\
		UCAC\,12065058 & 11.9 &5.2 &3.6 &$-$14.7 &5.2 &3.3 &85.0 & 0.2 & 9.5 &40/43/53/48 & 14.9 &14.1 &12.6 &12.1 &y* & 79\\
		UCAC\,12065064 & 1.3 &2.6 &0.3 &$-$4.2 &2.4 &0.2 &$-$14.5 & 0.6 & 0.5 &45/54/62/54 & 14.0 &13.5 &12.4 &12.1 &y & 3\\
		UCAC\,12065075 & $-$0.8 &2.6 &0.4 &1.5 &2.4 &2.1 &18.4 & 0.1 & 2.8 &45/52/63/58 & 14.1 &13.4 &12.1 &11.8 &y* & 10\\
		UCAC\,12284480 & 1.5 &5.2 &0.3 &$-$10.9 &5.2 &2.0 &17 & 3.0 & 2.7 &38/39/46/47 &  &  & 12.7 &11.9 &n & 14\\
		UCAC\,12284506 & $-$4.5 &2.6 &1.5 &$-$11.3 &2.4 &2.2 &$-$10 & 3.0 & 0.0 &43/54/66/62 & 13.9 &13.2 &11.8 &11.5 &y & 16\\
		UCAC\,12284534 & $-$2.4 &5.2 &0.9 &$-$18.8 &5.2 &4.7 &$-$70 & 3.0 & 6.0 &38/42/55/51 & 14.8 &13.9 &12.1 &11.6 &n & 55\\
		UCAC\,12284536 & 1.0 &2.5 &0.2 &$-$3.9 &2.4 &0.3 &$-$12.5 & 6.7 & 0.2 &63/77/93/74 & 12.8 &12.4 &11.5 &11.4 &y & 3\\
		UCAC\,12284546 & 1.9 &5.7 &0.5 &$-$3.4 &1.8 &0.5 &$-$16.1 & 2.3 & 0.6 &52/63/75/55 & 13.3 &12.9 &11.9 &11.6 &y & 4\\
		UCAC\,12284585 & $-$6.2 &1.4 &2.1 &$-$8.9 &1.4 &1.4 &51 & 3.0 & 6.1 &48/54/65/62 & 13.2 &12.5 &11.1 &10.6 &n & 20\\
		UCAC\,12284589 & $-$5.1 &3.0 &1.7 &$-$5.5 &1.8 &0.2 &$-$12 & 3.0 & 0.2 &76/101/104/103 & 12.1 &11.8 &11.2 &11.0 &y & 7\\
		UCAC\,12284594 & 0.8 &5.2 &0.1 &$-$3.1 &5.2 &0.6 &9.7 & 0.3 & 2.0 &40/42/56/48 & 14.6 &13.9 &12.5 &12.1 &y & 7\\
		UCAC\,12284608 & 3.5 &5.2 &1.0 &$-$31.0 &5.2 &8.7 &$-$55 & 3.0 & 4.5 &36/38/44/49 & 15.3 &14.3 &12.3 &11.6 &n & 100\\
		UCAC\,12284620 & $-$6.3 &5.2 &2.1 &$-$27.1 &5.2 &7.4 &$-$55 & 3.0 & 4.5 &41/46/56/55 & 14.3 &13.4 &11.8 &11.3 &n & 98\\
		UCAC\,12284626 & $-$11.5 &5.2 &3.7 &$-$28.6 &5.2 &7.9 &$-$15 & 3.0 & 0.5 &40/45/49/56 & 14.3 &13.5 &12.1 &11.7 &n & 100\\
		UCAC\,12284628 & 2.9 &1.4 &0.8 &$-$10.5 &3.4 &1.9 &$-$45.0 & 0.5 & 3.5 &53/72/80/66 & 13.0 &12.7 &11.8 &11.6 &y* & 14\\
		UCAC\,12284631 & 2.9 &1.4 &0.8 &$-$5.1 &1.4 &0.1 &$-$0.9 & 0.2 & 0.9 &62/81/97/79 & 11.8 &11.3 &11.3 &11.0 &y & 4\\
		UCAC\,12284638 & 1.0 &1.5 &0.2 &$-$1.4 &1.5 &1.1 &$-$6 & 3.0 & 0.4 &57/72/89/69 & 12.0 &11.8 &11.7 &11.4 &y & 3\\
		UCAC\,12284645 & 0.4 &1.4 &0.0 &$-$2.3 &1.4 &0.8 &$-$19.0 & 9.2 & 0.9 &65/76/96/72 & 12.8 &12.4 &11.5 &11.2 &y & 2\\
		UCAC\,12284653 & 3.3 &5.4 &0.9 &$-$3.3 &5.2 &0.5 &$-$13.4 & 0.6 & 0.3 &36/36/38/42 &  &  & 13.1 &12.6 &y & 8\\
		UCAC\,12284662 & 3.5 &5.3 &1.0 &$-$10.6 &5.2 &1.9 &$-$14.9 & 0.9 & 0.5 &44/46/56/53 & 14.4 &13.8 &12.3 &11.9 &y & 16\\
		UCAC\,12284746 & 0.1 &5.2 &0.1 &$-$4.3 &5.4 &0.2 &$-$10.9 & 0.2 & 0.1 &36/39/46/48 &  &  & 12.5 &12.2 &y & 7\\
		\hline
		\multicolumn{14}{l}{\textsuperscript{\phantom{*}*}\footnotesize{
		Star considered as members based only on photometry.}}
	\end{tabular}}
\end{table*}
\section{Observations of NGC\,6250}\label{Sect_Observations}
\subsection{Target}
The cluster NGC\,6250 is located in the constellation Ara in the
Southern hemisphere. Using proper motions and photometry from PPMXL
\citep{Roeser10} and 2MASS \citep{Skrutskie06} $JHK$ photometry data,
\citet{Kharchenko13} have estimated the age of the cluster to be
$\log t \sim 7.42$\,yr and the distance from the Sun to the
cluster as 865\,pc. \citet{Kharchenko13} calculated the cluster proper
motion as $0.7\,\pm\,0.4$\,mas\,yr$^{-1}$ in right ascension (RA) and
$-4.1\,\pm\,0.4$\,mas\,yr$^{-1}$ in declination (DEC) with a
cluster radial velocity of $-8.0\,\pm\,0.8$\,km\,s$^{-1}$. Previously,
\citet{Herbst77} estimated $\log t \sim 7.146$\,yr for the age and 
1025\,pc for the distance. \citet{Moffat75} estimated $d=950$\,pc.
NGC\,6250 is located in a dust-rich region of space, with $E$\,($B$\,--\,$V$) = 0.385 and
$E$\,($J$\,--\,$H$) = 0.123 \citep{Kharchenko13}. To our best knowledge, the cluster
has not previously been studied in detail spectroscopically, except  
for the purpose of classification spectra and radial velocity measurements.
\subsection{Instrument}
The observations of NGC\,6250 were obtained in service mode on 2007 May 27 and 
30 using FLAMES, the multi-object spectrograph attached 
to the Unit 2 Kueyen of the ESO/VLT.

The FLAMES instrument \citep{Pasquini02} is able to access targets
over a field of view of 25 arcmin in diameter. Its 138 fibres feed two
spectrographs, GIRAFFE and the Ultraviolet and Visual Echelle
Spectrograph (UVES). This makes it possible to observe up to 138
stars, 130 using GIRAFFE linked to FLAMES with MEDUSA fibres, and 8
using UVES linked to FLAMES with UVES fibres. FLAMES-GIRAFFE can
obtain low- or medium-resolution spectra ($R$ = 7500--30\,000), within the
spectral range 3700--9000\,\AA. Low-resolution spectra may be obtained
within wavelength intervals 500--1200\,\AA\ wide; medium-resolution
spectra are obtained in wavelength intervals 170--500\,\AA\ wide. FLAMES-UVES can 
obtain high-resolution spectra ($R$ = 47\,000),
with central wavelengths of 5200, 5800 or 8600\,\AA\, each covering a
wavelength range $\sim 2000$\,\AA\ wide.
\subsection{Instrument Settings}
For the observations, we chose the instrument set-up that allowed us:
(1) to observe two hydrogen lines, which are essential to the
determination of $T_{\rm eff}$ and $\log$ $g$ and (2) to maximize the
number of metal lines and consequently the number of chemical elements
available for spectral analysis.  The GIRAFFE settings were chosen
such that they are as close to the guiding wavelength of 520\,nm as
possible, in an effort to minimize light losses due to atmosphere
differential refraction. The higher spectral resolution of UVES means that,
to achieve a high enough S/N ratio for each spectra, UVES
requires an exposure time typically three or four times that of GIRAFFE. As a
result during UVES observations we are able to obtain observations
with three GIRAFFE settings. The settings we used are shown in Table
\ref{table:Settings}.  We observed with the HR9B and HR11 settings
both observing nights and one L- setting each night.

To prevent saturation of the observations each UVES observation was
divided into four sub-exposures and each GIRAFFE setting was divided
into two sub-exposures.
\subsection{Data reduction}
Our data were obtained in service mode, and the package released
included the products reduced by ESO with the instrument dedicated
pipelines. In this work, we used the low- and mid resolution GIRAFFE
data as reduced by ESO and re-reduced UVES spectra.

For the normalization we followed a standard procedure. First, we fit
the observed spectrum with a function $G(\lambda)$, clipping in an
iterative way all points 3\,$\sigma$ above and 1\,$\sigma$ below the
spectrum.  Then, we compare the normalized spectrum with a synthetic
spectrum calculated adopting a similar stellar model, in order to test
the quality of our normalization. Experience has shown for the low-resolution 
settings LR3 and LR6, a third-order polynomial gives the
best normalization. For UVES and the high-resolution settings HR9B and
HR11, a cubic spline gives the best normalization.  In the case of UVES
the Balmer lines spread across two different orders, the merging of
the orders was therefore done before the normalization.
\section{Cluster membership}\label{Sect_Membership}
\label{sec:clusmem}
\begin{figure}
\centering
\includegraphics[width=\columnwidth]{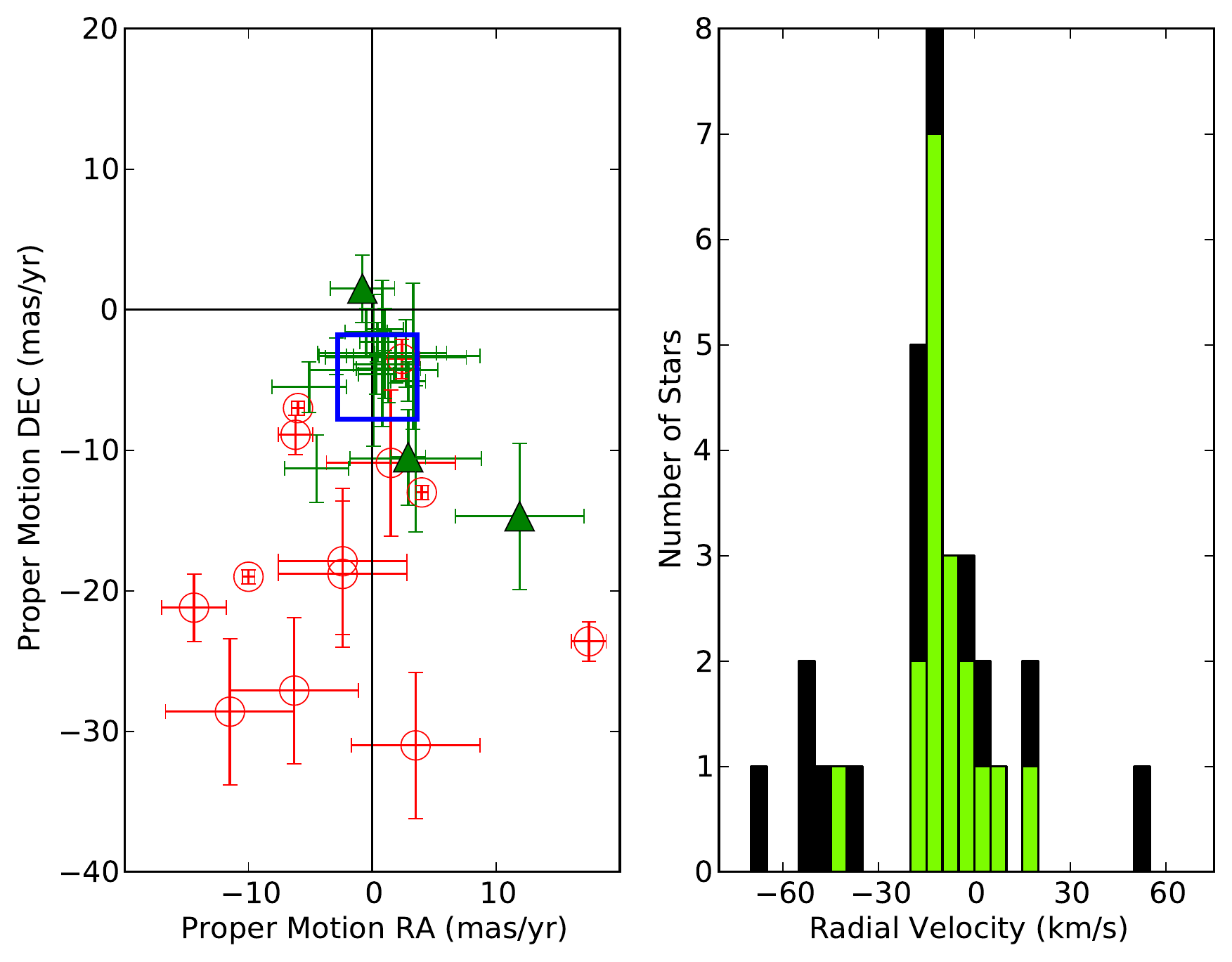}
\caption{Left-hand panel: the proper motion of the observed stars from
the UCAC2 catalogue \citep{Zacharias04}. Green plus-signs are stars we consider to be
members and red circles are stars we do not consider to be members. Triangles 
are the stars for which membership is considered based solely on 
photometry. The remaining points are stars for which membership is 
considered based on the kinematics and photometry. 
The blue box is centred at the cluster mean proper 
motions and represents the uncertainty of these values
 in each direction.  
Right-hand panel: histogram of the
 radial velocity measurements calculated using the method
described in Section \ref{sec:RadV} for non-member stars 
and for the member stars using {\sc sparti\_simple}. 
The green histogram highlights the radial velocity of the cluster 
members.}
\label{fig:Kine_mem}
\end{figure}
\begin{figure}
\centering
\includegraphics[width=\columnwidth]{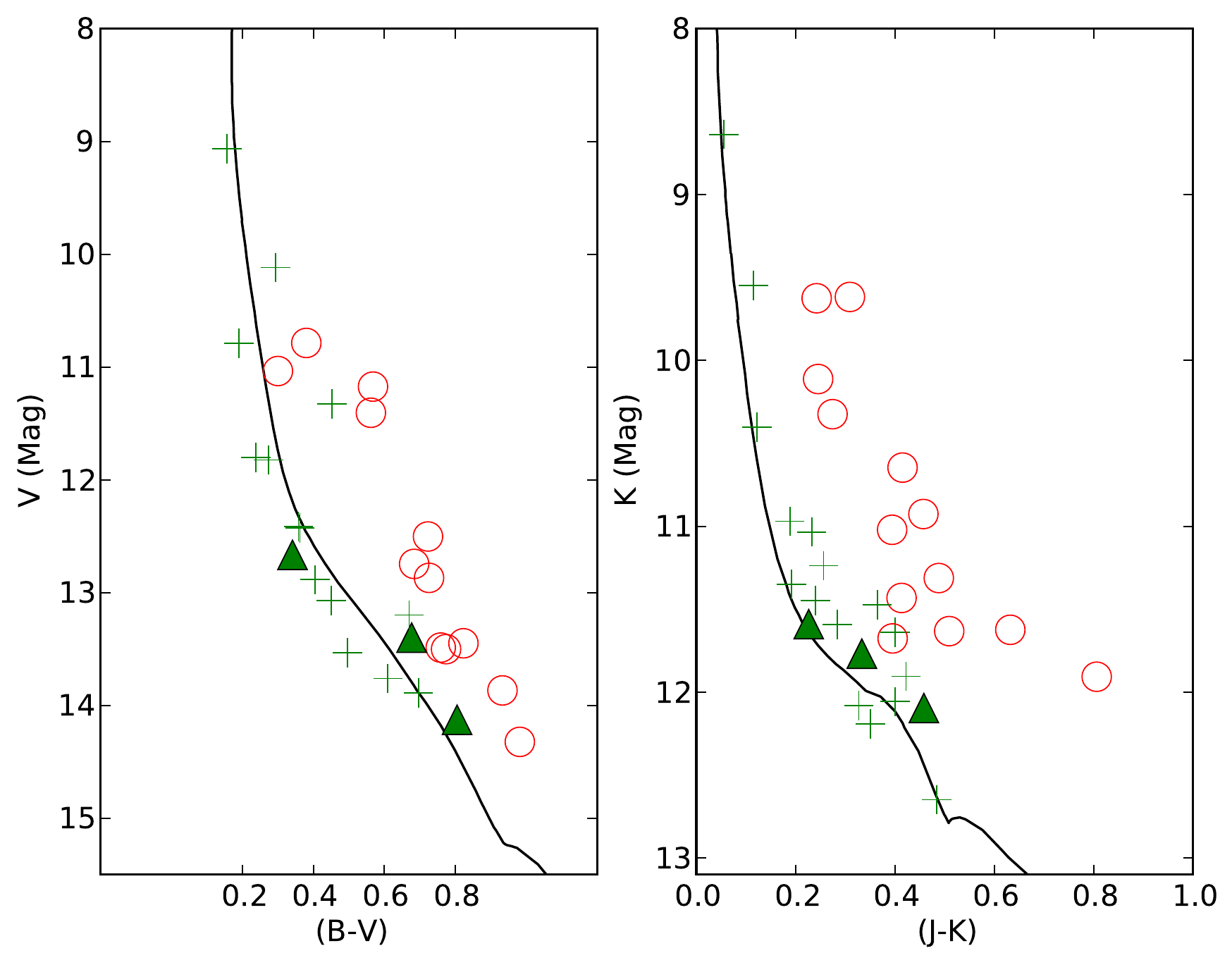}
\caption{
In both panels, green plus-signs are stars we consider
members and red circles are stars we do not consider to be members. Triangles 
are the stars for which membership is considered based solely on 
photometry. The remaining points are stars for which membership is 
considered based on the kinematics and photometry.
Left-hand panel: optical colour--magnitude diagram of the observed stars. The $BV$
photometry is taken from the APASS catalogue \citep{Henden2016}, 
plotted with the isochrone at ($\log t \sim 7.42$) corrected 
for $E$\,($B$\,--\,$V$) = 0.385 \citep[][black solid line]{Kharchenko13}. 
Right-hand panel: infrared colour--magnitude diagram of the observed stars. The $JK$
photometry is taken from the UCAC2 catalogue \citep{Zacharias04}, plotted with 
the isochrone at ($\log t \sim 7.42$) corrected for  $E$\,($J$\,--\,$K$) = 0.200  \citep[][black solid line]{Kharchenko13}.}
\label{fig:Phot_mem}
\end{figure}
\begin{figure*}
\centering
\includegraphics[width=.5\textwidth]{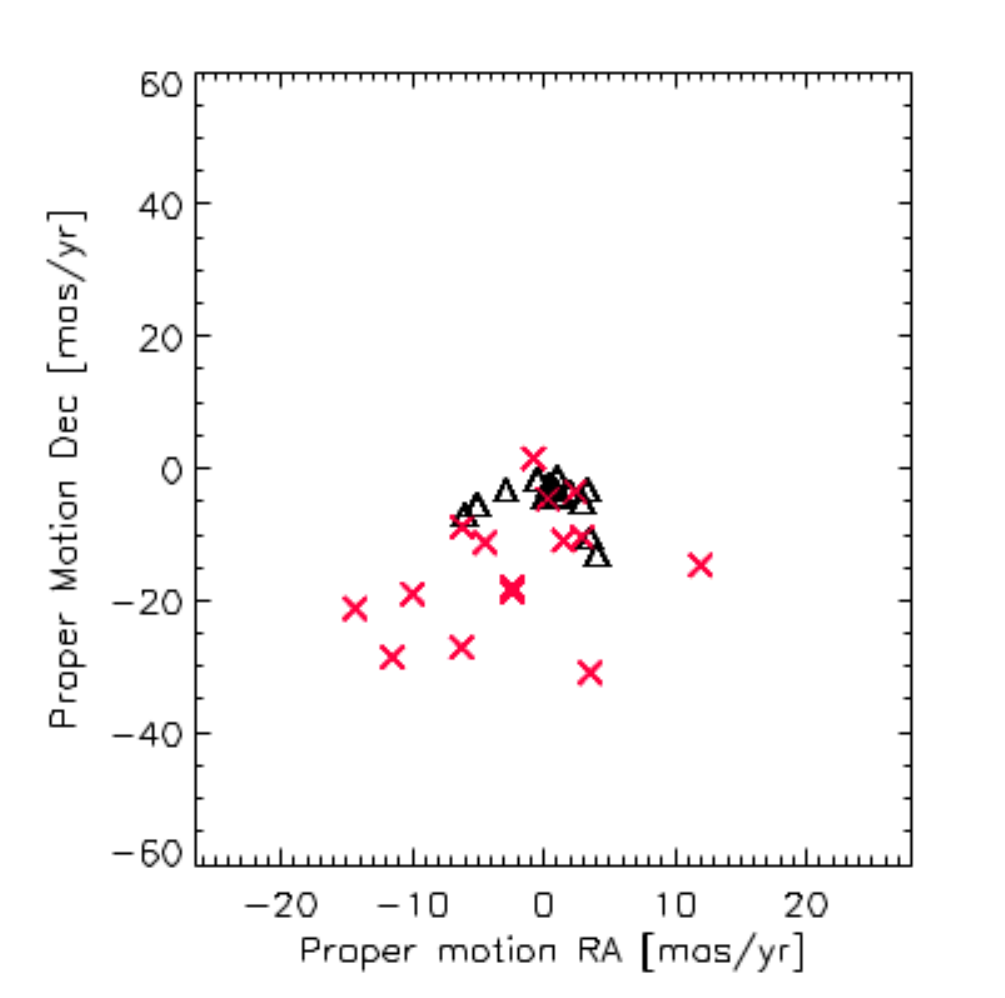}\includegraphics[width=.5\textwidth]{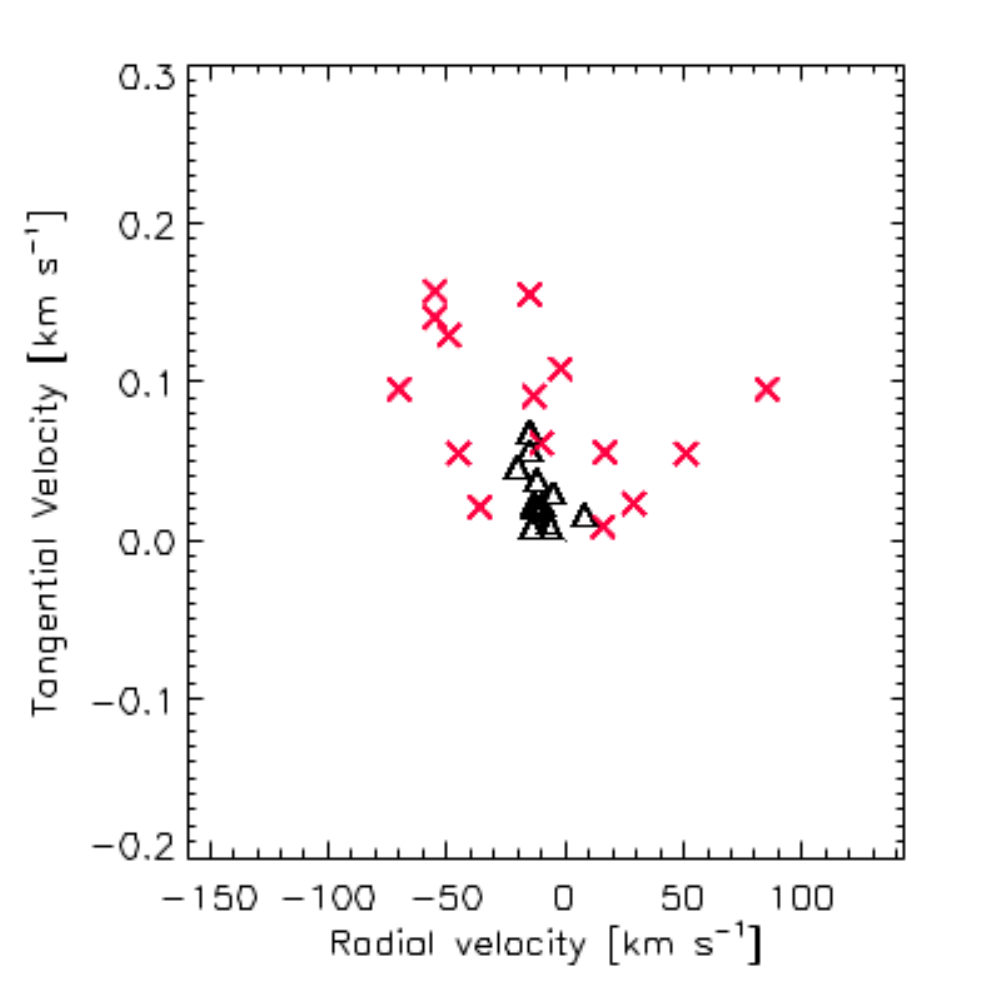}
\caption{$K$-means clustering result for the membership analysis of
  NGC6250: the black triangles are member stars and the red crosses are
  non-members. The left-hand panel shows the sample distribution in proper
  motion. The right-hand panel shows the sample distribution in
  velocity. The tangential velocity was computed using the
  distance 850\,pc given by \citet{Kharchenko13}}
\label{fig:CesMember}
\end{figure*}

Previous studies by \citet{Bayer00}, \citet{Dias06} and
\citet{Feinstein08} give cluster membership information for the stars
in the vicinity of NGC\,6250. \citet{Dias06} estimate cluster
membership using the statistical method of \citet{Sanders71}. The
probabilities are shown in Table \ref{Tab:Prog}. Note that some of the
results of \citet{Dias06} show inconsistencies: three stars
(UCAC\,12065057, UCAC\,12284608 and UCAC\,12284626) have been assigned
100\,\% likelihood of being members despite their proper motion values
being far from the cluster mean \citep[which they calculate to be consistent
with][]{Kharchenko13}. Furthermore, UCAC\,12065064 has proper
motion values very close to the cluster mean values and its photometry 
fits well with the theoretical isochrone, it only has a
3\,\% likelihood of being a member.

The determination of cluster membership from previous literature was
used as a guess for an initial target selection. Our new spectroscopic
data allow a refined membership study, which is critical for our analysis.
\subsection{Proper motion and radial velocity membership} 
The proper motions and radial velocities of the observed stars are
shown in Table~\ref{Tab:Prog} and are plotted in 
Fig.~\ref{fig:Kine_mem}. In our membership analysis we have
followed two methods.

First, we have identified those stars that are within 1\,$\sigma$ of
the mean proper motion and radial velocity values for our sample.
This way we have identified 19 stars to be members of the
cluster. We calculate the cluster mean proper motions as,
0.1\,$\pm$\,2.9\,mas\,yr$^{-1}$ in RA,
$-$6.1\,$\pm$\,4.4\,mas\,yr$^{-1}$ in DEC and radial velocity of
$-$10\,$\pm$\,11\,km\,s$^{-1}$.

As a cross-check, we also use the partitional clustering
technique, $K$-means clustering \citep{Macqueen1967}. It is a technique to 
find common data points based on the analysis of the variables 
that define the data. In general, the method 
is to set a predicted number of  data clusters and give an initial guess 
for the centres of these clusters. The algorithm then assigns points 
to the closest cluster centre and recalculates until the 
cluster centre values do not change. Our problem is simplified
by the fact that we can define one cluster centre close
to the literature value of cluster proper motion and radial
velocity. In practice, we have set the number of clusters to 5: one
initially centred in the literature values $ (\mu_0, v_{r_0}) $ and
the other four initially centred respectively at $(\mu_0 +
1.5\,\sigma_\mu, v_{r_0})$, $(\mu_0 - 1.5\,\sigma_\mu, v_{r_0})$,
$(\mu_0, v_{r_0} + 1.5\,\sigma_{v_{r}})$ and $(\mu_0, v_{r_0} +
1.5\,\sigma_{v_r})$, where $\sigma_{v_{r}}$ and $\sigma_\mu$ are the
standard deviation in our sample. We have performed this computation
using the CLUSTER function in {\sc idl}.

Fig.~\ref{fig:CesMember} shows the results of our cluster analysis. We
found 15 members and we calculated the cluster mean proper motions
as 0.4\,$\pm$\,3.0\,mas\,yr$^{-1}$ in RA,
$-$4.80\,$\pm$\,3.2\,mas\,yr$^{-1}$ in DEC and radial velocity of
$-$10\,$\pm$\,6\,km\,s$^{-1}$. Both methods give the same results
apart from four stars.  The discrepancy between the two methods is likely
because the spread of radial velocity is large and asymmetric, and the
$K$-means clustering is able to deal with this more effectively.

\subsection{Photometry}

 The magnitude and colour of our sample of stars are a good 
indicator of cluster membership. Using the $B$ and $V$ magnitudes 
from  \citet{Henden2016} and the $J$ and $K$ 
magnitudes from \citet{Zacharias04}, we have produced two colour magnitude
diagrams, displayed in Fig.~\ref{fig:Phot_mem}. Each diagram shows the photometry 
and a theoretical isochrone calculated with CMD 2.7 \citep{Bressan2012,Chen2014,Tang2014,Chen2015} 
for an age of $\log t = 7.42$. The photometry has been corrected for 
the extinction \citep{Kharchenko13} and distance to the cluster 
\citep[850 pc;][]{Kharchenko13}.

 In general, we see very good agreement between the member stars 
 determined using the kinematics approach and those that fit the 
 isochrones. Notable exceptions are UCAC\,12065058, UCAC\,12065075 and 
 UCAC\,12284628, which 
do not agree with the kinematics of the cluster mean, but agree very 
well with the photometry. This maybe as a result of a re-ejection 
event and so we consider these stars as members. 
\section{Fundamental parameters}\label{Sect_Fundamental}
\subsection{Radial velocity and rotational velocity}
\label{sec:RadV}

We have calculated initial values of $v_{\rm rad}$ and \vsini\ from the
FLAMES spectra obtained in the range 5150--5350\,\AA\ with the
highest resolution setting HR9B. We used a least-squares
deconvolution \citep[LSD;][]{Kochukhov2010} of the observed spectrum with lines
selected from the VALD list \citep{Piskunov95} to calculate an average line profile 
in velocity space. To do the line selection,
we must have an estimate of the temperature, so we use a combination
of Balmer lines and photometry to determine an estimate of the temperature.
\begin{figure}
\centering
\includegraphics[width=\columnwidth]{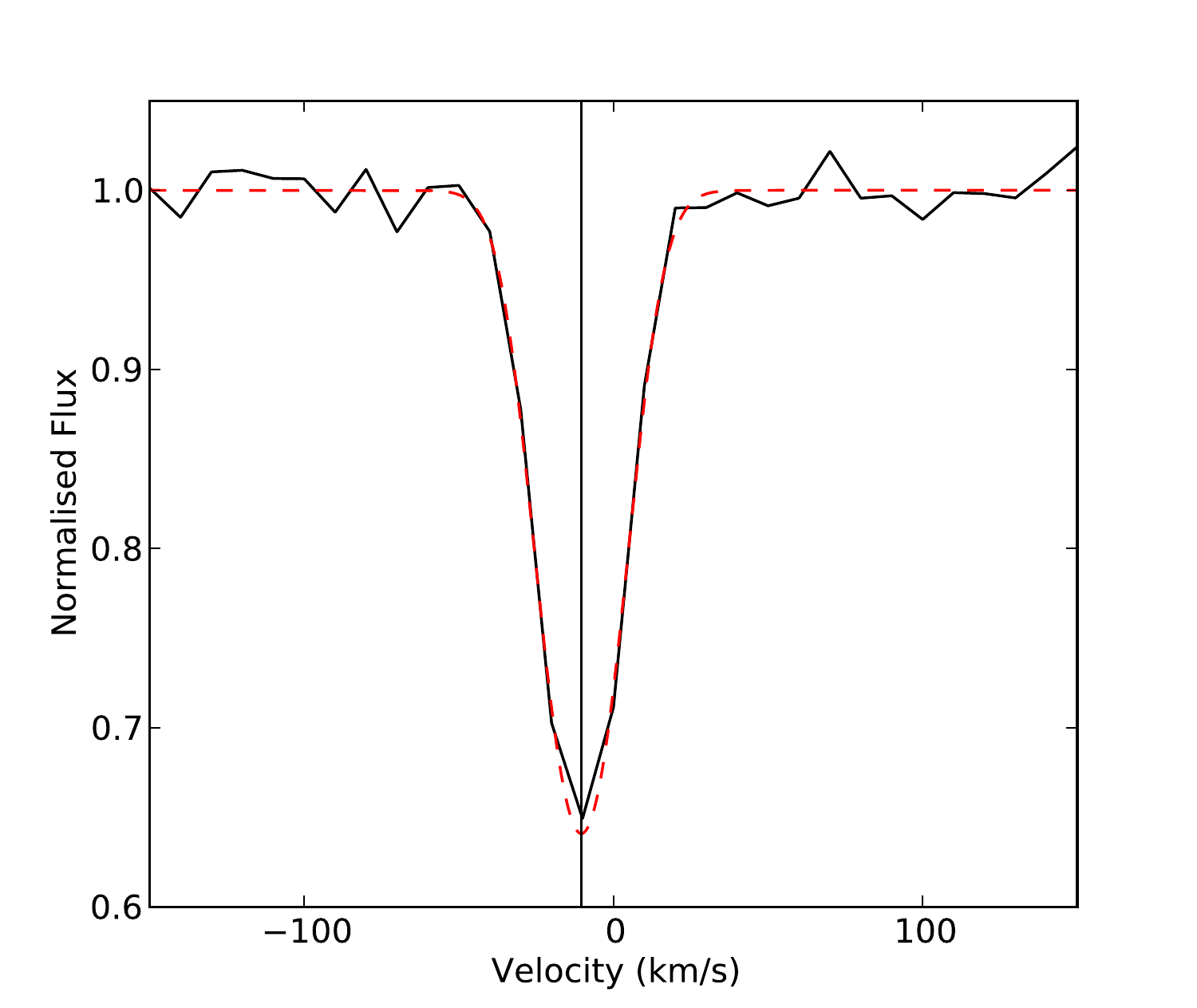}
\caption{The least-square deconvolution of the spectrum of
  UCAC\,12284746 ($T_{\rm eff} = 7200$\,K) (solid black line), plotted against
  the Gaussian fit (red dashed line). The radial velocity is given
  by the position of the centre of the Gaussian, and it is 
  indicated by a solid vertical line.}
\label{fig:lsd}
\end{figure}
\begin{figure}
\centering
\includegraphics[width=\columnwidth]{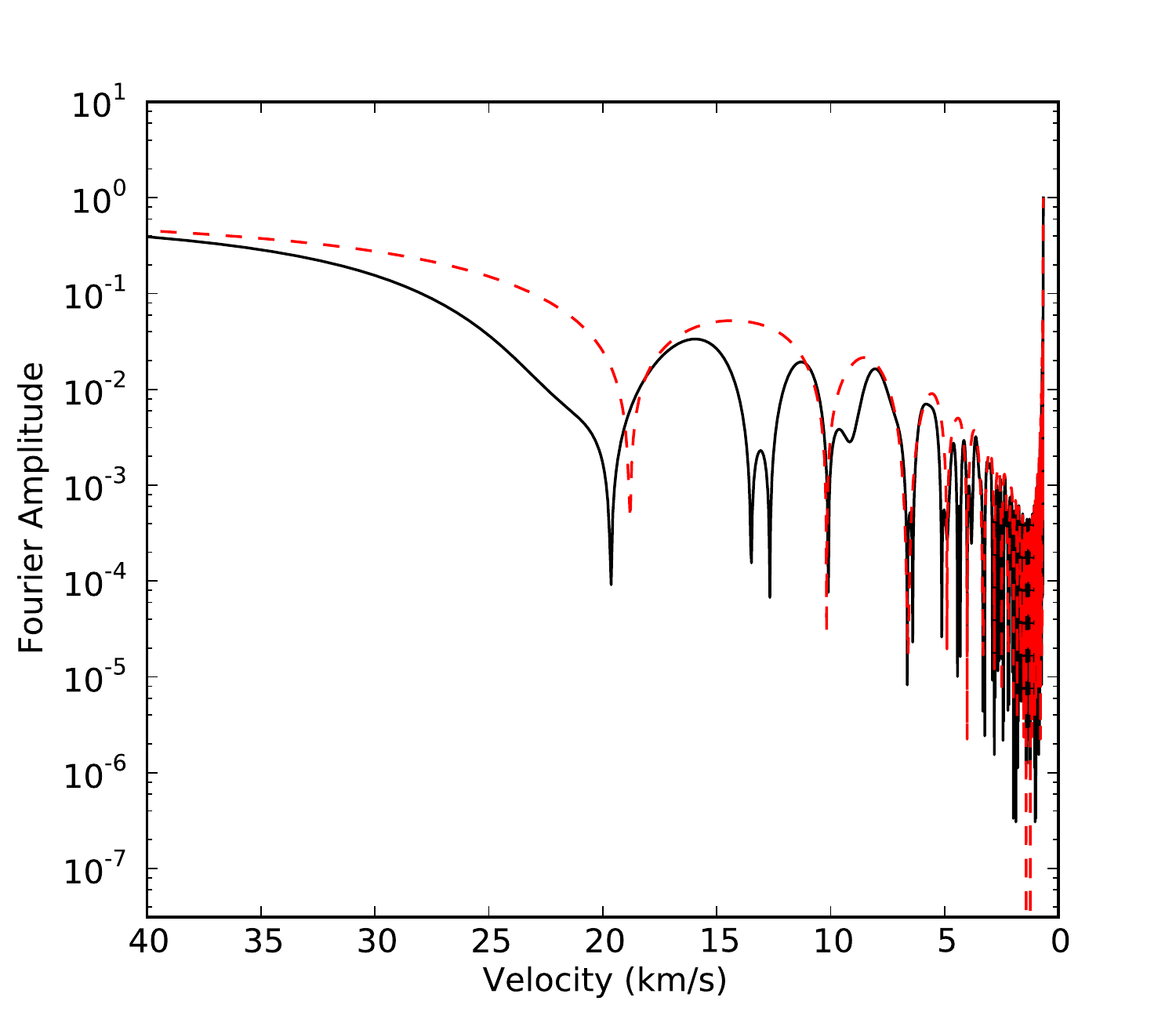}
\caption{The fast Fourier transform of the LSD profile
  of Fig. \ref{fig:lsd} (black solid line), plotted with the
  FTT of a model LSD profile with
  $\teff = 7200$\,K and $\vsini =19.5$\,\kms\ (red dashed lines).}
\label{fig:vsini}
\end{figure}
  The radial velocity of each star has been estimated using a Gaussian fit, as shown in Fig.~\ref{fig:lsd}.

In order to measure \vsini\, the LSD profile is then shifted to the rest frame, 
and the fast Fourier transform (FFT) is calculated. An example of the FFT is 
shown in Fig.~\ref{fig:vsini}. Following \citet{Gray2005}, the first minimum of the FFT corresponds 
to the stellar \vsini\ value. \citet{Glazunova2008} showed that it is possible 
to use the LSD profile, in place of the more noisy profiles of single lines, to 
derive the \vsini\ value using the FFT method described by \citet{Gray2005}.
Because we expect low $v_{\rm mac}$ values  \citep{Grassitelli2015}, at the resolution of FLAMES 
$v_{\rm mac}$ cannot be distinguished from $v \sin i$ and we therefore ignore it.
\subsection{Fundamental parameters from Balmer lines}
\begin{figure}
\centering
   \includegraphics[width=\columnwidth]{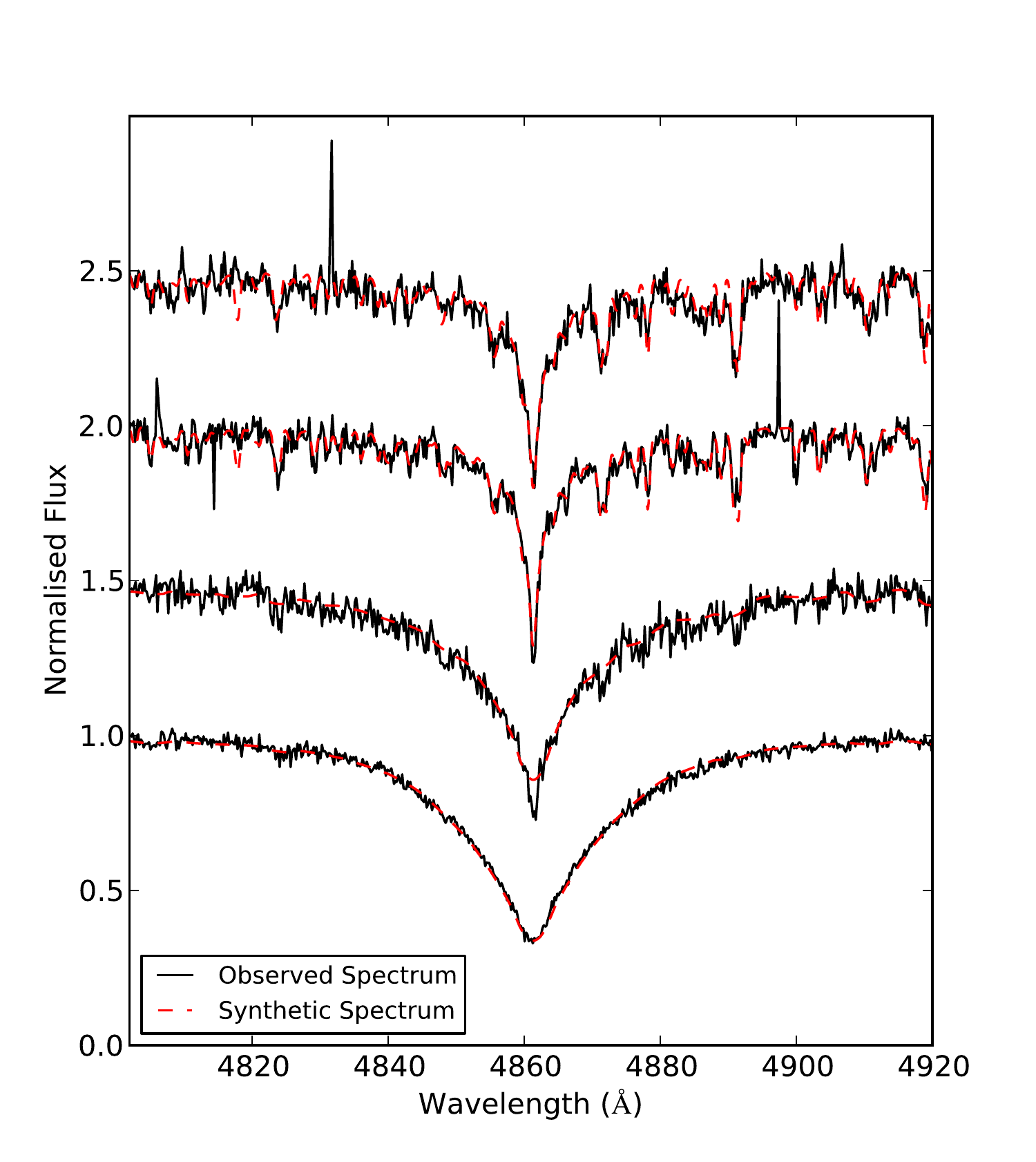}
     \caption{A sample of the observed H\,$\beta$ lines (black solid lines) fitted
     with the model spectra (red dashed line). From top to bottom
     the stars are:  
     UCAC\,12284594     ($T_{\rm eff} = 6200$\,K), 
     UCAC\,12065075         ($T_{\rm eff} = 6300$\,K), 
     UCAC\,12065064     ($T_{\rm eff} = 7600$\,K) and
     UCAC\,12284536 ($T_{\rm eff} = 9800$\,K). Each profile is calculated 
     with $v\sin i$ as shown in Table~\ref{table:AtomPars}.}
     \label{fig:balmer}
\end{figure}

We deduced the fundamental parameters \teff\ and \logg\ by fitting
synthetic and observed Balmer lines H\,$\beta$ and H\,$\gamma$ for UVES
spectra and H\,$\alpha$ and H\,$\beta$ for GIRAFFE spectra.  Model
atmospheres were computed with \atlasnine\ \citep{Kurucz93} assuming
plane parallel geometry, local thermodynamical equilibrium and opacity
distribution function (ODF) for solar abundances
\citep{Kurucz93b}. The synthetic spectra were computed with
\cossamsimplex. Examples of the fit between our model and the observed Balmer lines are shown in Fig. \ref{fig:balmer}.

We used hydrogen line as both temperature and gravity indicator
because for $\teff \la 8000$\,K they are more sensitive to temperature
and for higher temperature they are more sensitive to $\log g$
variation but temperature effects can still be visible in the part of
the wing close to the line core, according to \citet{Fossati11a}.\\

\begin{figure}
\centering
   \includegraphics[width=\columnwidth]{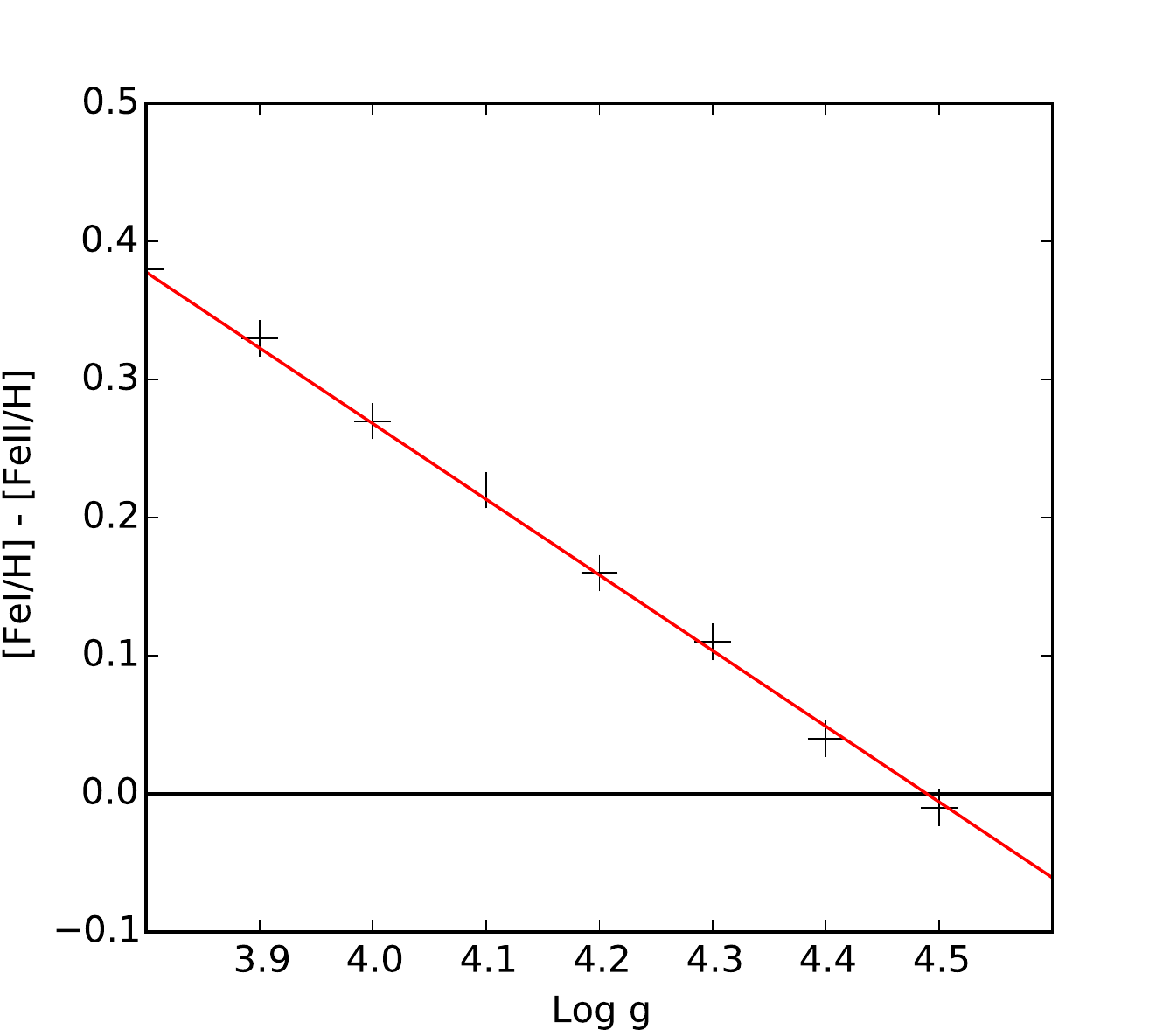}
     \caption{The difference between the abundance of Fe {\sc i} and Fe {\sc ii} 
     plotted as a function of surface gravity for  the Sun determined using  \spartisimplex\ 
     by varying the \logg\ of the model atmosphere at $T_{\rm eff}$ = 5800K.}
     \label{fig:IonBal}
\end{figure}

\begin{figure}
\centering
   \includegraphics[width=\columnwidth]{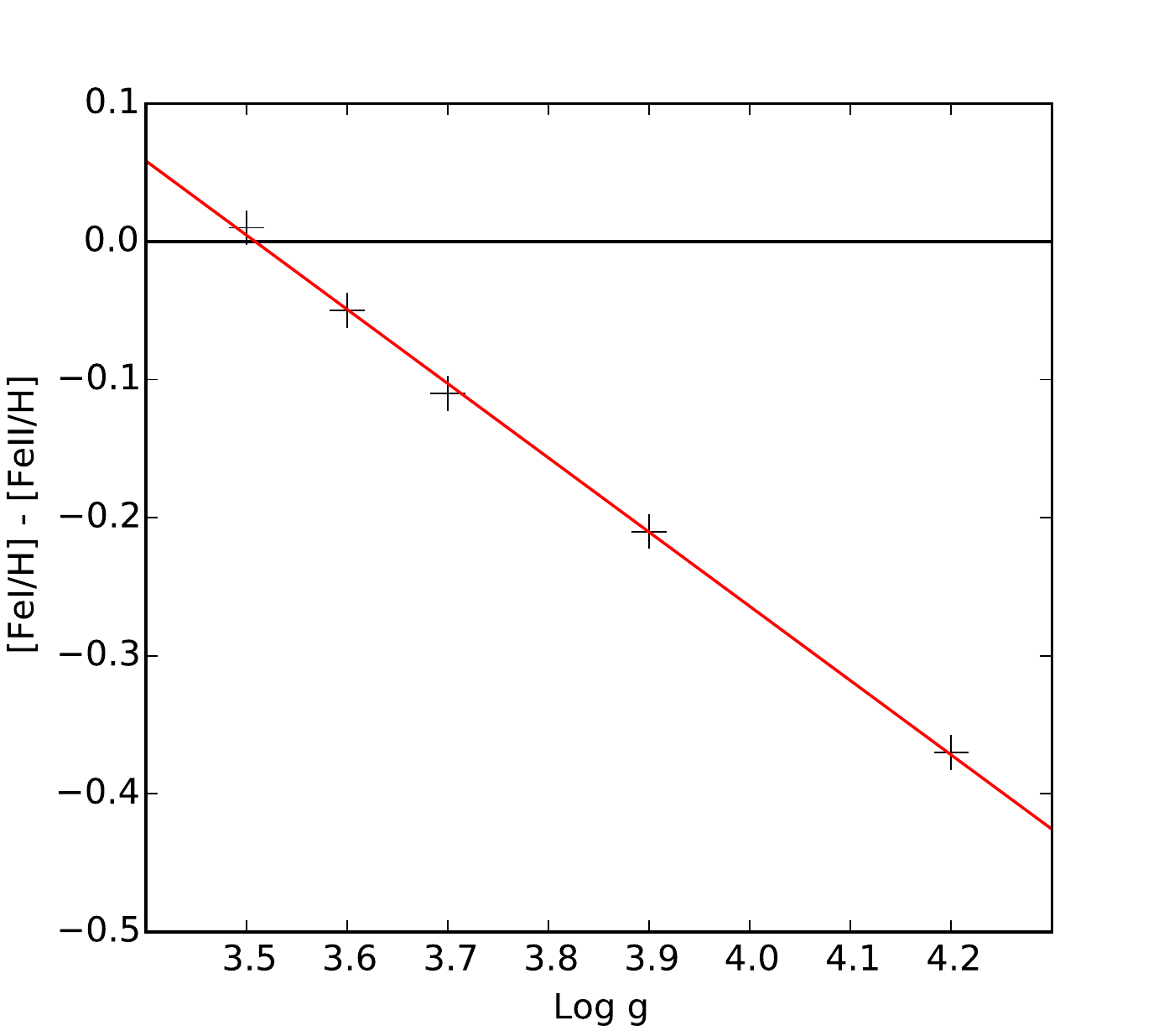}
     \caption{The difference between the abundance of Fe {\sc i} and Fe {\sc ii} 
     plotted as a function of surface gravity for  the star 21Peg determined using  \spartisimplex\ 
     by varying the \logg\ of the model atmosphere at $T_{\rm eff}$ = 10400K.}
     \label{fig:IonBal21Peg}
\end{figure}

To check our values of \logg\ for each star we determined the \logg\
which provided the best ionization balance between Fe {\sc i} and Fe {\sc ii} lines. 
To test this method, we calculate the abundance of Fe {\sc i} and Fe {\sc ii} for 
the Sun using \teff\ = 5800\,K and varying the $\log g$ between  3.8 
and 4.5. The value of $\log g$ where the abundances of Fe {\sc i} and Fe {\sc ii} are equal 
is 4.49 compared with 4.44 found by \citep{Prsa2016}. We performed  a similar analysis for 21 Peg, varying the $\log g$ between 3.5 and 4.2 and using $T_{\rm eff}$ = 10400\,K. As a result we calculate a value for $\log g$ of 3.5 compared with 3.55 found by \citet{Fossati09}. The results of this analysis are shown in Figs \ref{fig:IonBal} and Figs \ref{fig:IonBal21Peg}.
For each of the NGC\,6250 stars, the $\log g$ values found using 
both methods agree within the uncertainties.

As a result of the low S/N, we were unable to measure any abundances for 
TYC\,8327-565-1 and UCAC\,12284638; however, we were able to estimate 
$T_{\rm eff}$ and $\log g$ that are given in Table \ref{table:AtomPars}.

\begin{table}
	\caption{ Atmospheric parameters for the sample of stars from NGC\,6250.} 
	\label{table:AtomPars}     
	\centering
	\resizebox{\columnwidth}{!}{\begin{tabular}{l r@{ $\pm$ }c r@{ $\pm$ }c r@{ $\pm$ }c r@{ $\pm$ }c}
		\hline\hline\\[-2.0ex]
		Star & 
		\multicolumn{2}{c}{$T_{\rm eff}$} & 
		\multicolumn{2}{c}{$\log g$}    &
		\multicolumn{2}{c}{$v_{\rm mic}$} &
		\multicolumn{2}{c}{$v \sin i$} \\
		& 
		\multicolumn{2}{c}{(K)} & 
		\multicolumn{2}{c}{(CGS)} &
		\multicolumn{2}{c}{(\kms)} &
		\multicolumn{2}{c}{(\kms)} \\
		\hline\\[-2.0ex]   
		HD\,152706	&  9900	&  200 & 4.2	& 0.1& 1.9	& 0.1 & 151.6 & 0.2\\
		HD\,152743	& 19\,800	&  200 & 4.1	& 0.1& 2.1	& 2.5 & 198.9 & 12.7\\
		NGC\,6250-13	&  8200	&  200 & 4.2	& 0.1& 2.0	& 0.1 & 70.1 & 0.8\\
		TYC\,8327-565-1	& 14\,200	&  500 & 4.2	& 0.3 & \multicolumn{2}{c}{--} &\multicolumn{2}{c}{--}\\
		UCAC\,12065058	&  6000	&  200 & 4.4	& 0.1& 0.9	& 0.2 & 4.8 & 0.9\\
		UCAC\,12065064	&  7600	&  200 & 4.2	& 0.1& 0.3	& 0.4 & 54.1 & 0.7\\
		UCAC\,12065075	&  6300	&  200 & 4.4	& 0.1& 1.0	& 0.1 & 15.2 & 0.2\\
		UCAC\,12284506	&  6100	&  200 & 4.4	& 0.1& 0.6	& 0.1 & 10.6 & 0.2\\
		UCAC\,12284536	& 10\,000	&  200 & 4.4	& 0.1& 2.0	& 0.2 & 170.2 & 3.6\\
		UCAC\,12284546	&  8400	&  200 & 4.2	& 0.1& 0.3	& 0.5 & 146.0 & 1.5\\
		UCAC\,12284589	& 12\,600	&  200 & 4.2	& 0.1& 0.5	& 0.7 & 215.5 & 2.7\\
		UCAC\,12284594	&  6100	&  200 & 4.4	& 0.1& 0.3	& 0.2 & 50.3 & 0.5\\
		UCAC\,12284628	& 10\,000	&  200 & 4.3	& 0.1& 0.3	& 0.5 & 22.9 & 0.8\\
		UCAC\,12284631	&  9800	&  200 & 4.2	& 0.1& 0.7	& 0.2 & 17.8 & 0.4\\
		UCAC\,12284638	& 10\,800	&  400 & 4.2	& 0.3&  \multicolumn{2}{c}{--} &\multicolumn{2}{c}{--}\\
		UCAC\,12284645	& 11\,000	&  200 & 4.3	& 0.1& 2.0	& 0.4 & 270.0 & 3.0\\
		UCAC\,12284653	&  6200	&  200 & 4.4	& 0.1& 1.5	& 0.2 & 49.5 & 1.0\\
		UCAC\,12284662	&  7400	&  200 & 4.4	& 0.1& 2.0	& 0.1 & 81.4 & 0.9\\
		UCAC\,12284746	&  7200	&  200 & 4.3	& 0.1& 1.8	& 0.2 & 24.5 & 0.4\\

		\hline\\[-2.0ex]
	\end{tabular}}
\end{table}
\begin{table*}
	\label{table:Abuns}
	\caption{ The abundance of elements for the analysed stars of NGC6250 (ordered by decreasing $T_{\rm eff}$), given in log($N$/H) 
	where H = 12.00. In parenthesis, the first number is the error calculated using equation (\ref{eq:cov})  and the 
	second in the error calculating using equation (\ref{eq:err}). Both errors are in units of 0.01dex. The last row 
	of each set gives the solar abundances from \citet{Asplund09}.}   
	\label{table:results}     
	\centering
	\resizebox*{!}{\dimexpr\textheight-2\baselineskip\relax}{\begin{tabular}{l r r@{ $\pm$ }r c c c c c c}	
		\hline\hline\\[-2.0ex]
		Star & $T_{\rm eff}$ & \multicolumn{2}{c}{$v \sin i$} & C & O & Na & Mg & Si & S\\
             & (K)                & \multicolumn{2}{c}{(km\,s$^{-1}$)}\\
		\hline\\[-2.0ex]
		HD\,152743	& 19\,800 & 198.9 & 12.7	&		&		&		&		& 8.07\,(95;102) 	&	\\
		UCAC\,12284589	& 12\,600 & 215.5 & 2.7	&		&		&		&		&		&	\\
		UCAC\,12284645	& 11\,000 & 270.0 & 3.0	&		&		&		&		&		&	\\
		UCAC\,12284628	& 10\,000 & 22.9 & 0.8	&		&		&		& 6.41\,(06;12) 	&		&	\\
		UCAC\,12284536	& 10\,000 & 170.2 & 3.5	&		&		&		& 7.58\,(25;27) 	&		&	\\
		HD\,152706	&  9900 & 151.6 & 0.2	& 8.91\,(02;34) 	& 8.68\,(02;04) 	&		& 7.99\,(02;28) 	& 7.83\,(02;27) 	&	\\
		UCAC\,12284631	&  9800 & 17.8 & 0.4	& 8.84\,(13;16) 	&		&		& 6.61\,(04;12) 	& 7.98\,(11;27) 	&	\\
		UCAC\,12284546	&  8400 & 146.0 & 1.5	& 8.83\,(09;15) 	&		&		& 6.19\,(15;67) 	&		&	\\
		NGC\,6250-13	&  8200 & 70.1 & 0.8	&		&		&		& 7.74\,(06;14) 	&		&	\\
		UCAC\,12065064	&  7600 & 54.1 & 0.7	& 8.63\,(11;15) 	&		&		& 7.29\,(08;13) 	& 7.39\,(08;15) 	& 7.92\,(11;59) \\
		UCAC\,12284662	&  7400 & 81.4 & 0.9	& 8.50\,(19;37) 	&		& 6.24\,(17;20) 	& 7.59\,(09;13) 	& 7.51\,(10;10) 	&	\\
		UCAC\,12284746	&  7200 & 24.5 & 0.4	& 8.92\,(13;14) 	&		& 6.25\,(13;15) 	& 7.68\,(10;15) 	& 7.26\,(10;11) 	&	\\
		UCAC\,12065075	&  6300 & 15.2 & 0.2	&		&		& 6.49\,(07;22) 	& 7.64\,(04;22) 	& 7.34\,(04;19) 	&	\\
		UCAC\,12284653	&  6200 & 49.5 & 1.0	&		&		& 6.37\,(21;26) 	& 7.58\,(11;19) 	& 7.50\,(13;14) 	&	\\
		UCAC\,12284594	&  6100 & 50.3 & 0.5	&		&		& 6.24\,(09;10) 	& 7.45\,(03;12) 	& 7.11\,(06;07) 	&	\\
		UCAC\,12284506	&  6100 & 10.6 & 0.2	& 8.91\,(08;26) 	&		& 6.05\,(05;07) 	& 7.28\,(01;10) 	& 6.94\,(03;04) 	& 7.81\,(28;46) \\
		UCAC\,12065058	&  6000 & 4.8 & 0.9	&		&		& 6.08\,(17;17) 	& 7.70\,(08;12) 	& 6.91\,(12;12) 	&	\\
		{\bf Solar} & {\bf 5777} & \multicolumn{2}{c}{\bf 1.2} 	& {\bf 8.43}	& {\bf 8.69}	& {\bf 6.24}	& {\bf 7.60}	& {\bf 7.51}	& {\bf 7.12}\\
 		\hline\\[-2.0ex]
		Star & $T_{\rm eff}$ & \multicolumn{2}{c}{$v \sin i$} & Ca & Sc & Ti & V & Cr & Mn\\
             & (K)               & \multicolumn{2}{c}{(km\,s$^{-1}$)}\\
		\hline\\[-2.0ex]
		HD\,152743	& 19\,800 & 198.9 & 12.7	&		&		&		&		&		&	\\
		UCAC\,12284589	& 12\,600 & 215.5 & 2.7	&		&		&		&		& 6.25\,(14;18) 	&	\\
		UCAC\,12284645	& 11\,000 & 270.0 & 3.0	&		&		& 4.96\,(28;35) 	&		& 5.64\,(23;32) 	&	\\
		UCAC\,12284628	& 10\,000 & 22.9 & 0.8	&		&		& 4.94\,(11;16) 	&		& 5.57\,(09;10) 	&	\\
		UCAC\,12284536	& 10\,000 & 170.2 & 3.5	&		&		& 5.02\,(23;23) 	&		& 5.73\,(21;23) 	&	\\
		HD\,152706	&  9900 & 151.6 & 0.2	& 6.92\,(02;46) 	&		& 4.95\,(02;04) 	&		& 5.78\,(02;16) 	&	\\
		UCAC\,12284631	&  9800 & 17.8 & 0.4	&		&		& 4.75\,(06;08) 	&		& 5.49\,(07;08) 	&	\\
		UCAC\,12284546	&  8400 & 146.0 & 1.5	& 7.14\,(22;22) 	&		& 5.05\,(09;19) 	&		& 6.14\,(07;20) 	&	\\
		NGC\,6250-13	&  8200 & 70.1 & 0.8	& 6.22\,(08;14) 	&		& 4.83\,(05;08) 	&		& 5.75\,(06;09) 	& 5.51\,(13;17) \\
		UCAC\,12065064	&  7600 & 54.1 & 0.7	& 6.63\,(17;24) 	&		& 5.24\,(02;08) 	&		& 5.95\,(07;12) 	& 5.63\,(23;35) \\
		UCAC\,12284662	&  7400 & 81.4 & 0.9	& 6.32\,(12;21) 	& 3.13\,(14;23) 	& 4.97\,(06;14) 	&		& 5.69\,(08;23) 	& 5.45\,(16;21) \\
		UCAC\,12284746	&  7200 & 24.5 & 0.4	& 6.19\,(13;16) 	& 3.11\,(11;12) 	& 5.23\,(06;10) 	&		& 5.92\,(07;10) 	& 5.46\,(16;18) \\
		UCAC\,12065075	&  6300 & 15.2 & 0.2	& 6.55\,(07;21) 	& 3.41\,(05;11) 	& 5.12\,(03;13) 	& 4.44\,(07;18) 	& 5.92\,(03;17) 	& 5.27\,(10;15) \\
		UCAC\,12284653	&  6200 & 49.5 & 1.0	& 6.40\,(23;28) 	& 2.95\,(24;26) 	& 5.19\,(09;42) 	& 3.99\,(34;55) 	& 6.06\,(09;25) 	& 5.62\,(23;47) \\
		UCAC\,12284594	&  6100 & 50.3 & 0.5	& 6.33\,(09;11) 	& 2.90\,(13;14) 	& 5.00\,(04;05) 	& 4.78\,(07;09) 	& 5.92\,(03;08) 	& 5.18\,(15;17) \\
		UCAC\,12284506	&  6100 & 10.6 & 0.2	& 6.22\,(06;08) 	& 3.16\,(05;06) 	& 4.93\,(02;05) 	& 3.81\,(07;11) 	& 5.53\,(02;07) 	& 5.30\,(07;09) \\
		UCAC\,12065058	&  6000 & 4.8 & 0.9	& 6.42\,(16;16) 	& 3.01\,(15;16) 	& 4.92\,(08;10) 	& 4.05\,(18;20) 	& 5.62\,(07;10) 	& 5.25\,(22;23) \\
		{\bf Solar} & {\bf 5777} & \multicolumn{2}{c}{\bf 1.2} 	& {\bf 6.34}	& {\bf 3.15}	& {\bf 4.95}	& {\bf 3.93}	& {\bf 5.64}	& {\bf 5.43}\\
 		\hline\\[-2.0ex]
		Star & $T_{\rm eff}$ & \multicolumn{2}{c}{$v \sin i$} & Fe & Ni & Zn & Y & Ba & Nd\\
             & (K)                & \multicolumn{2}{c}{(km\,s$^{-1}$)}\\
		\hline\\[-2.0ex]
		HD\,152743	& 19\,800 & 198.9 & 12.7	& 7.67\,(47;49) 	&		&		&		&		&	\\
		UCAC\,12284589	& 12\,600 & 215.5 & 2.7	& 7.77\,(04;04) 	&		&		&		&		&	\\
		UCAC\,12284645	& 11\,000 & 270.0 & 3.0	& 7.50\,(12;23) 	&		&		&		&		&	\\
		UCAC\,12284628	& 10\,000 & 22.9 & 0.8	& 7.43\,(06;10) 	&		&		&		&		&	\\
		UCAC\,12284536	& 10\,000 & 170.2 & 3.5	& 7.56\,(16;17) 	&		&		&		&		&	\\
		HD\,152706	&  9900 & 151.6 & 0.2	& 7.61\,(01;12) 	&		&		&		&		&	\\
		UCAC\,12284631	&  9800 & 17.8 & 0.4	& 7.41\,(03;06) 	&		&		&		&		&	\\
		UCAC\,12284546	&  8400 & 146.0 & 1.5	& 7.72\,(07;23) 	& 6.94\,(10;17) 	&		&		&		&	\\
		NGC\,6250-13	&  8200 & 70.1 & 0.8	& 7.54\,(04;08) 	& 6.33\,(08;13) 	&		&		& 2.09\,(42;45) 	&	\\
		UCAC\,12065064	&  7600 & 54.1 & 0.7	& 7.74\,(05;14) 	& 6.16\,(11;13) 	&		&		&		&	\\
		UCAC\,12284662	&  7400 & 81.4 & 0.9	& 7.51\,(07;10) 	& 6.27\,(10;28) 	&		&		& 2.34\,(36;38) 	&	\\
		UCAC\,12284746	&  7200 & 24.5 & 0.4	& 7.73\,(07;11) 	& 6.30\,(08;12) 	& 4.71\,(44;46) 	& 2.54\,(13;14) 	& 2.18\,(49;49) 	&	\\
		UCAC\,12065075	&  6300 & 15.2 & 0.2	& 7.65\,(03;19) 	& 6.28\,(04;14) 	& 4.06\,(33;34) 	& 2.38\,(11;14) 	& 2.18\,(17;17) 	& 1.76\,(09;12) \\
		UCAC\,12284653	&  6200 & 49.5 & 1.0	& 7.70\,(09;20) 	& 6.44\,(11;27) 	& 4.62\,(88;88) 	&		& 2.20\,(41;43) 	&	\\
		UCAC\,12284594	&  6100 & 50.3 & 0.5	& 7.54\,(02;09) 	& 6.15\,(06;07) 	&		&		& 2.41\,(13;14) 	& 1.98\,(14;17) \\
		UCAC\,12284506	&  6100 & 10.6 & 0.2	& 7.28\,(01;07) 	& 5.99\,(03;05) 	& 4.69\,(28;28) 	& 2.07\,(09;11) 	& 2.42\,(09;09) 	& 1.61\,(07;13) \\
		UCAC\,12065058	&  6000 & 4.8 & 0.9	& 7.39\,(06;09) 	& 6.02\,(10;11) 	&		&		& 2.18\,(39;41) 	& 1.63\,(25;25) \\
		{\bf Solar} & {\bf 5777} & \multicolumn{2}{c}{\bf 1.2} 	& {\bf 7.50}	& {\bf 6.22}	& {\bf 4.56}	& {\bf 2.21}	& {\bf 2.18}	& {\bf 1.42}\\
 		\hline\\[-2.0ex]
\end{tabular}}
\end{table*}
\section{Results and Discussion}\label{Sect_Results}
The results of the abundance analysis are given in Table \ref{table:Abuns}. 
Since this is a young cluster, there is the potential for some of the stars 
to still have discs. If discs were present, we would expect to see the 
presence of emission lines, particularly in the core of H\,$\alpha$ and H\,$\beta$. 
We do not see any evidence of emission lines in any of the stars.

\subsection{UCAC\,12284546}
UCAC\,12284546 shows an overabundance of C, Ca, Cr, Fe and Ni 
and an underabundance of Mg. However, this abundance pattern 
does not match any 
standard chemically peculiar star in this temperature range. To better 
understand this star, it would be necessary to collect and analyse a 
higher resolution spectrum with higher S/N. As a result of the abundance 
anomalies we observe in this star, we do not consider this star in the 
global analysis of the results.
\subsection{Stellar Metallicity}
For the evolutionary tracks and isochrones we adopted the metallicity calculated as
\begin{equation}
Z_{\rm cluster} = 10^{[{\rm Fe/H}]_{\rm stars} - [{\rm Fe/H}]_{\rm \odot}}Z_{\rm \odot},
\label{eq:zphy} 
\end{equation}
where $Z_{\rm cluster}$ and $[{\rm Fe/H}]_{\rm stars}$ are respectively the clusters metallicity and average Fe abundance. 
This formulation does not follow the definition of $Z$, which is
\begin{equation}
Z=\sum_{i=1}^{n}m_iX_i,
\label{eq:zFe} 
\end{equation}
where $n$ is the number of elements, $m_i$ is the atomic mass of each element and $X_i$ the abundance of each element. In equation (\ref{eq:zFe}), $Z$ is driven mostly by the abundance of C and O, which are the most abundant elements following H and He, but in stellar evolutionary calculations the relevant factor is the Fe opacity. This is why for the cluster metallicity we adopt the expression given by equation (\ref{eq:zphy}). When using equation (\ref{eq:zphy}) to infer the metallicity, it is important to use as Z$_{\rm \odot}$ the value adopted by the considered stellar evolution tracks. In this work we use the stellar evolutionary tracks by \citet{Bressan2012}, which adopt Z$_{\rm \odot} = 0.0152$. Using the average Fe abundance obtained from the non-chemically peculiar stars, we obtain $Z_{\rm cluster} = 0.018\pm 0.005$, which is consistent with the solar value within the uncertainty.
\subsection{Spectroscopic H-R diagram}
\begin{figure}
\centering
   \includegraphics[width=\columnwidth]{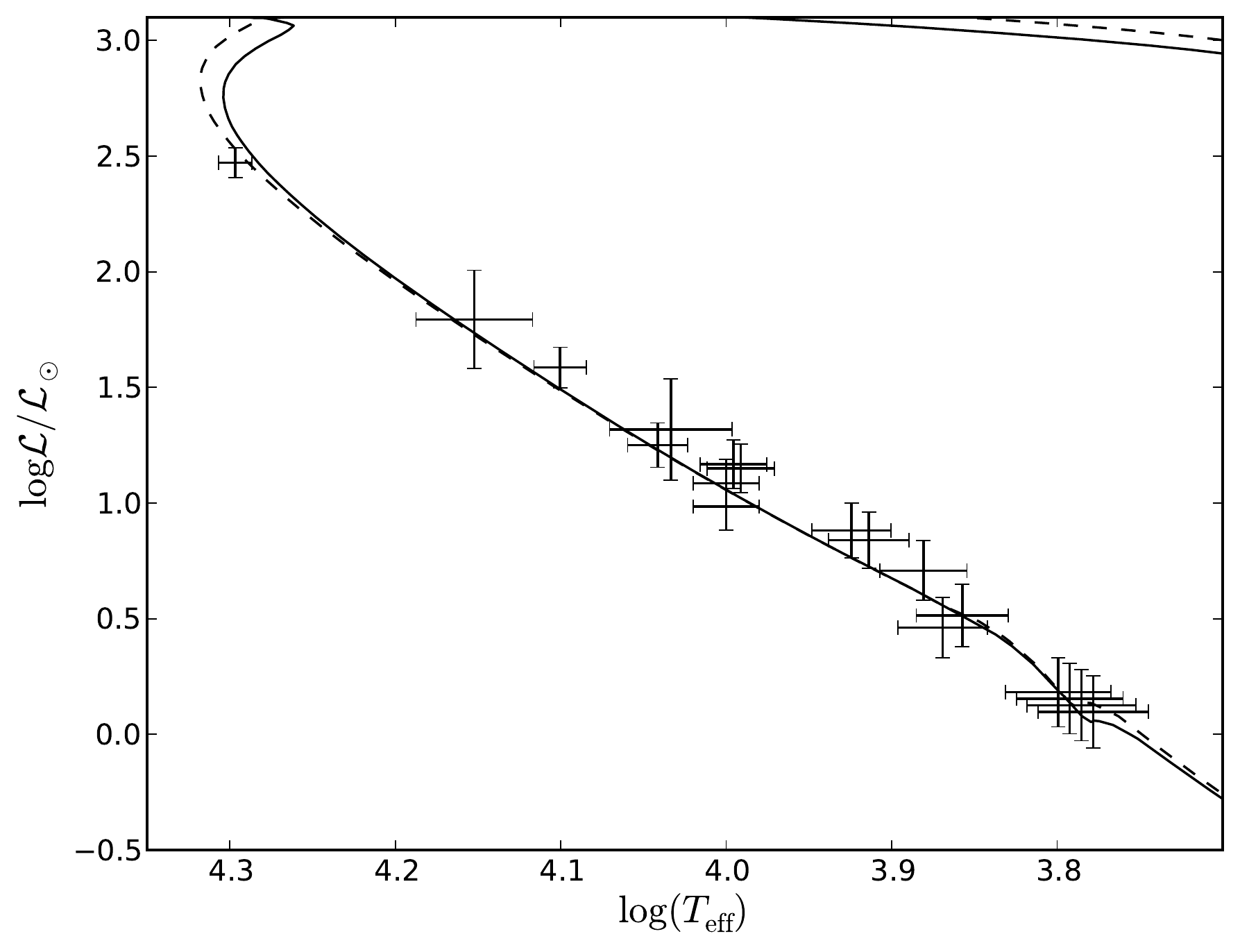}
     \caption{ An H-R diagram of NGC6250, the stars plotted (black plus-signs)
     with theoretical isochrones \citep{Bressan2012} at $\log t$ = 7.40 (dashed black line) and  
     $\log t$ = 7.45 (solid black line). Both isochrones have solar metallicity.}
     \label{fig:HD}
\end{figure}

We plot a spectroscopic H-R diagram (Fig. \ref{fig:HD}) using the \teff\ and \logg\ 
values calculated for each star. We calculate 
the flux weighted luminosity, $\log \mathcal{L}/\mathcal{L}_\odot$,  following \citet{Langer2014} with
\begin{equation}
\log \mathcal{L}/\mathcal{L}{\rm \odot} = 	\log\left(\frac{T_{\rm eff}^4}{g}\right)- 
								\log\left(\frac{T_{\rm eff\odot}^4}{g_\odot}\right)
\end{equation}
where \teff\ and \logg\ are taken from Table \ref{table:AtomPars}, $T_{\rm eff\odot}$ is the solar 
effective temperature and $g_\odot$ is the solar surface gravity. The isochrones are from 
\citet{Bressan2012}. 
Based on the H-R diagram, we are not able to constrain the age of this 
cluster; however, the age  
of $\log t$ = 7.42 given by \citet{Kharchenko13} fits well our data. As a 
result, we use this age in the remainder of the paper.
We also give the flux-weighted luminosity, masses and fractional age of each of the stars in 
Table \ref{table:isochrones}     
calculated by fitting evolutionary tracks \citep{Bressan2012} to each star. 
\begin{table}
	\caption{ $\log \mathcal{L}/\mathcal{L}_\odot$, $\log T_{\rm eff}$, $M/M_\odot$ and fractional age ($\tau$)  with associated
error bars for the stars of the NGC 6250 open cluster.}
	\label{table:isochrones}     
	\centering
	\resizebox{\columnwidth}{!}{\begin{tabular}{l r@{ $\pm$ }c r@{ $\pm$ }c  r@{ $\pm$ }c r@{ $\pm$ }c}
		\hline\hline\\[-2.0ex]
		Star & 
		\multicolumn{2}{c}{$\log \mathcal{L}/\mathcal{L}_\odot$} &
		\multicolumn{2}{c}{$\log T_{\rm eff}$}   &
		\multicolumn{2}{c}{$M/M_\odot$} &
		\multicolumn{2}{c}{$\tau$} \\
		\hline\\[-2.0ex]   
		HD\,152706	& 1.17	& 0.10 & 4.00 & 0.02	& 2.25	& 1.00	& 0.02	& 0.04\\
		HD\,152743	& 2.47	& 0.06 & 4.30 & 0.01	& 6.40	& 0.60	& 0.44	& 0.09\\
		NGC\,6250-13	& 0.84	& 0.12 & 3.91 & 0.02	& 1.75	& 1.00	& 0.01	& 0.03\\
		TYC\,8327-565-1	& 1.79	& 0.21 & 4.15 & 0.04	& 3.80	& 0.20	& 0.12	& 0.14\\
		UCAC\,12065058	& 0.10	& 0.16 & 3.78 & 0.03	& 1.05	& 0.10	& -0.02	& 0.01\\
		UCAC\,12065064	& 0.71	& 0.13 & 3.88 & 0.03	& 1.60	& 0.05	& 0.00	& 0.02\\
		UCAC\,12065075	& 0.18	& 0.15 & 3.80 & 0.03	& 1.15	& 0.05	& -0.01	& 0.02\\
		UCAC\,12284506	& 0.13	& 0.15 & 3.79 & 0.03	& 1.05	& 0.10	& -0.02	& 0.01\\
		UCAC\,12284536	& 0.98	& 0.10 & 4.00 & 0.02	& 2.05	& 0.10	& 0.02	& 0.04\\
		UCAC\,12284546	& 0.88	& 0.12 & 3.92 & 0.02	& 1.80	& 1.00	& 0.01	& 0.03\\
		UCAC\,12284589	& 1.59	& 0.09 & 4.10 & 0.02	& 3.20	& 1.00	& 0.07	& 0.01\\
		UCAC\,12284594	& 0.13	& 0.15 & 3.79 & 0.03	& 1.05	& 0.10	& -0.02	& 0.01\\
		UCAC\,12284628	& 1.08	& 0.10 & 4.00 & 0.02	& 2.20	& 0.10	& 0.02	& 0.04\\
		UCAC\,12284631	& 1.15	& 0.11 & 3.99 & 0.02	& 2.20	& 1.00	& 0.02	& 0.04\\
		UCAC\,12284638	& 1.32	& 0.22 & 4.03 & 0.04	& 2.60	& 0.20	& 0.04	& 0.06\\
		UCAC\,12284645	& 1.25	& 0.10 & 4.04 & 0.02	& 2.40	& 0.20	& 0.03	& 0.05\\
		UCAC\,12284653	& 0.15	& 0.15 & 3.79 & 0.03	& 1.10	& 0.05	& -0.02	& 0.01\\
		UCAC\,12284662	& 0.46	& 0.13 & 3.87 & 0.03	& 1.40	& 0.10	& -0.00	& 0.02\\
		UCAC\,12284746	& 0.51	& 0.13 & 3.86 & 0.03	& 1.45	& 0.05	& -0.00	& 0.02\\
		\hline\\[-2.0ex]
	\end{tabular}}
\end{table}

\begin{figure}
\centering
   \includegraphics[width=\columnwidth]{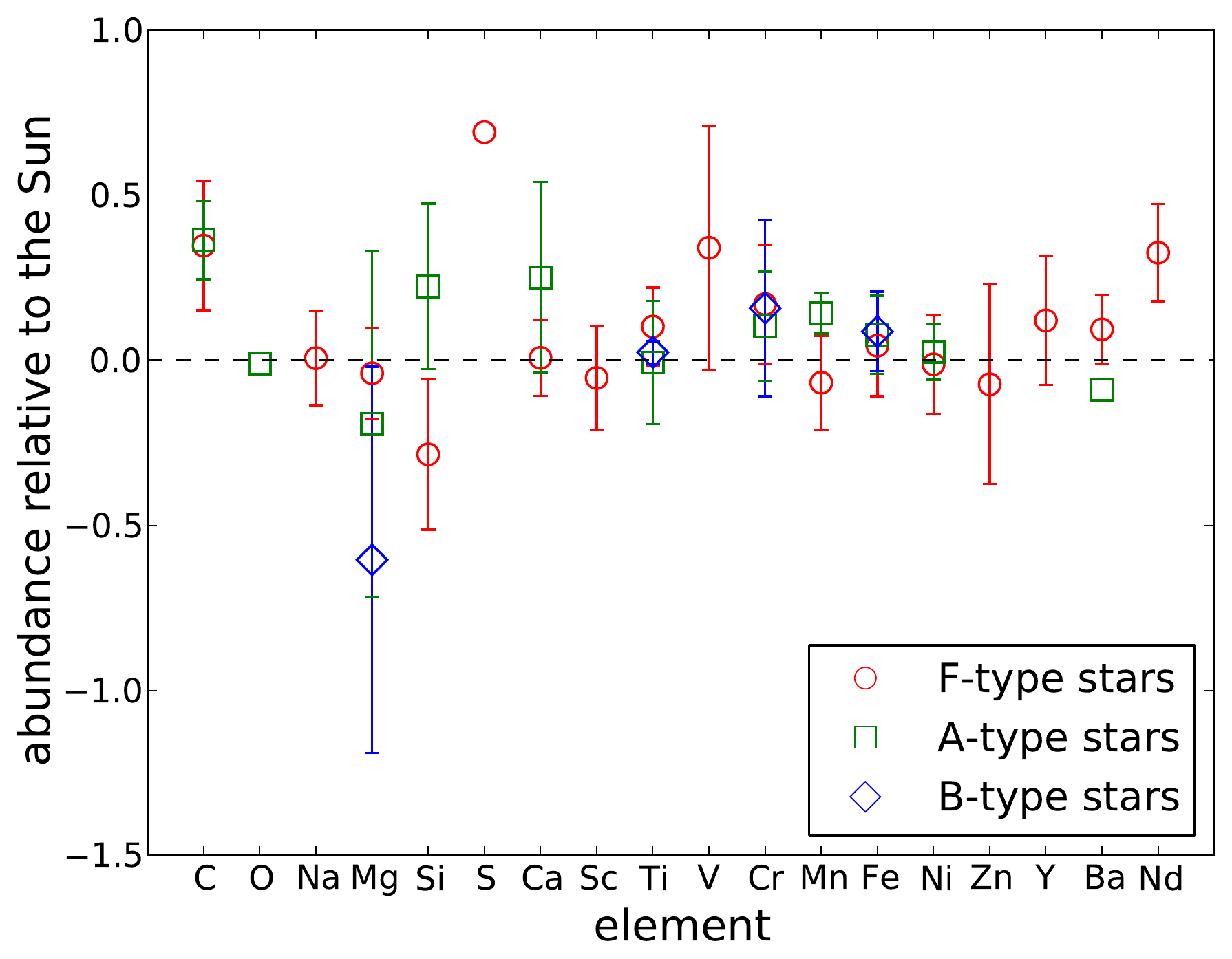}
     \caption{ The mean abundances of each element relative to solar 
     for F- (red circles), 
     A-(green squares) and B- (blue diamonds) type stars. The error bars
     are calculated by taking the standard deviations of the calculated mean
     abundances.}
     \label{fig:solar}
\end{figure}
\begin{figure}
\centering
   \includegraphics[width=\columnwidth]{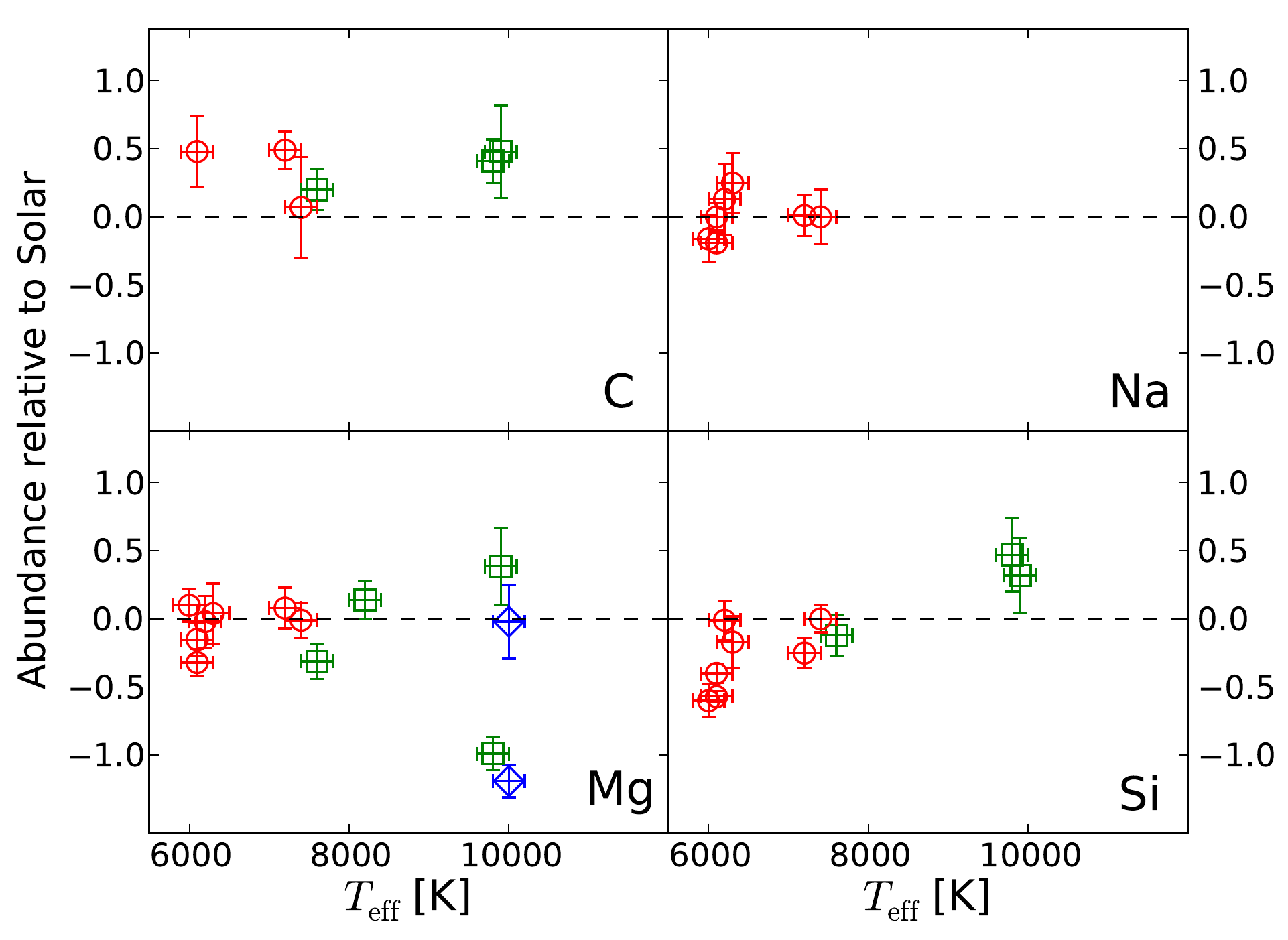}
     \caption{The abundances of C, Na, Mg, and Si relative to the solar
     abundance \citep{Asplund09} against $T_{\rm eff}$. For F- (red circles), 
     A-(green squares) and B- (blue diamonds) type stars.}
     \label{fig:Teff1}
\end{figure}
\begin{figure}
\centering
   \includegraphics[width=\columnwidth]{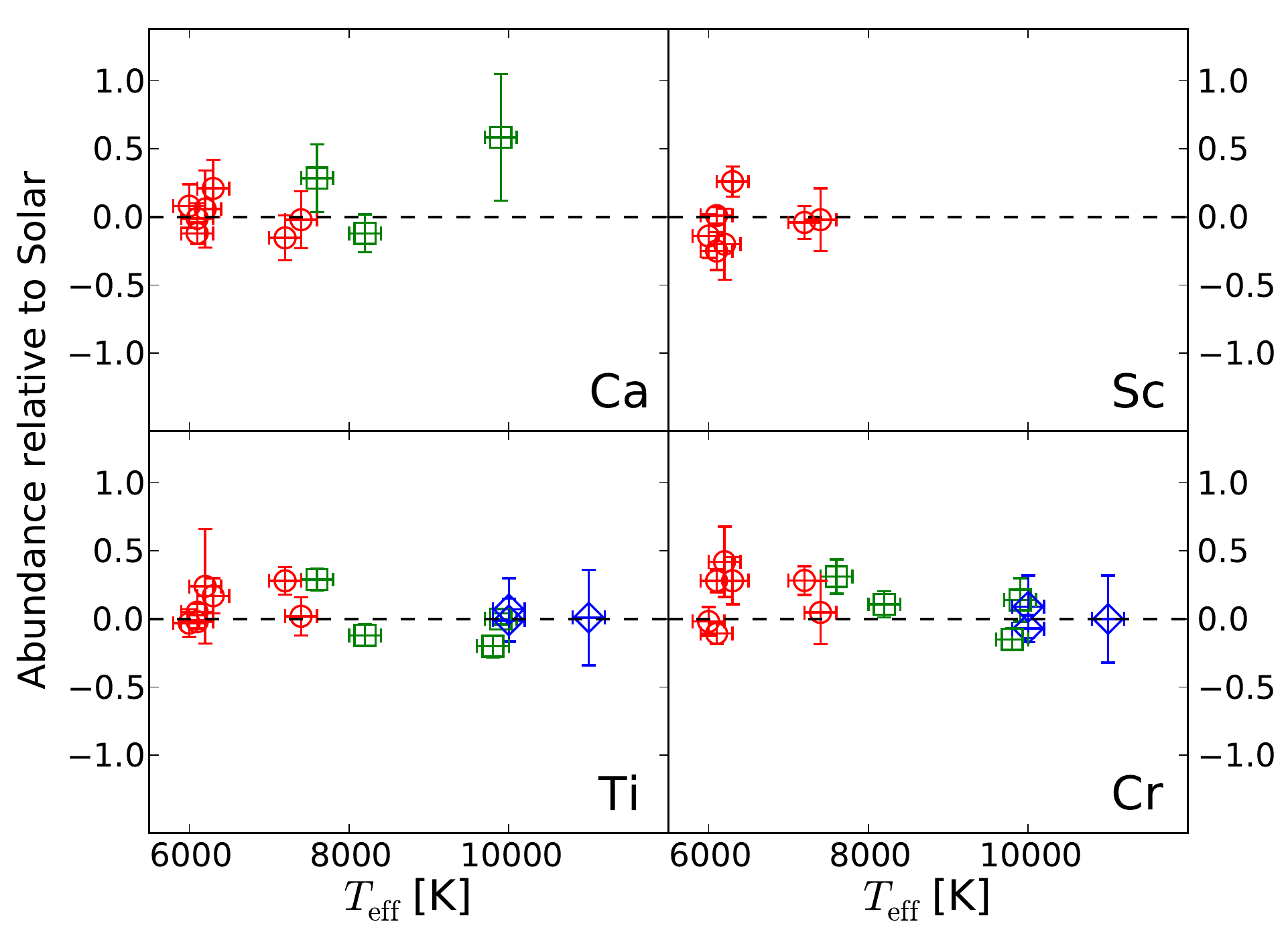}
     \caption{Same as Fig. \ref{fig:Teff1}, but for Ca, Sc, Ti and
     Cr.}
     \label{fig:Teff2}
\end{figure}
\begin{figure}
\centering
   \includegraphics[width=\columnwidth]{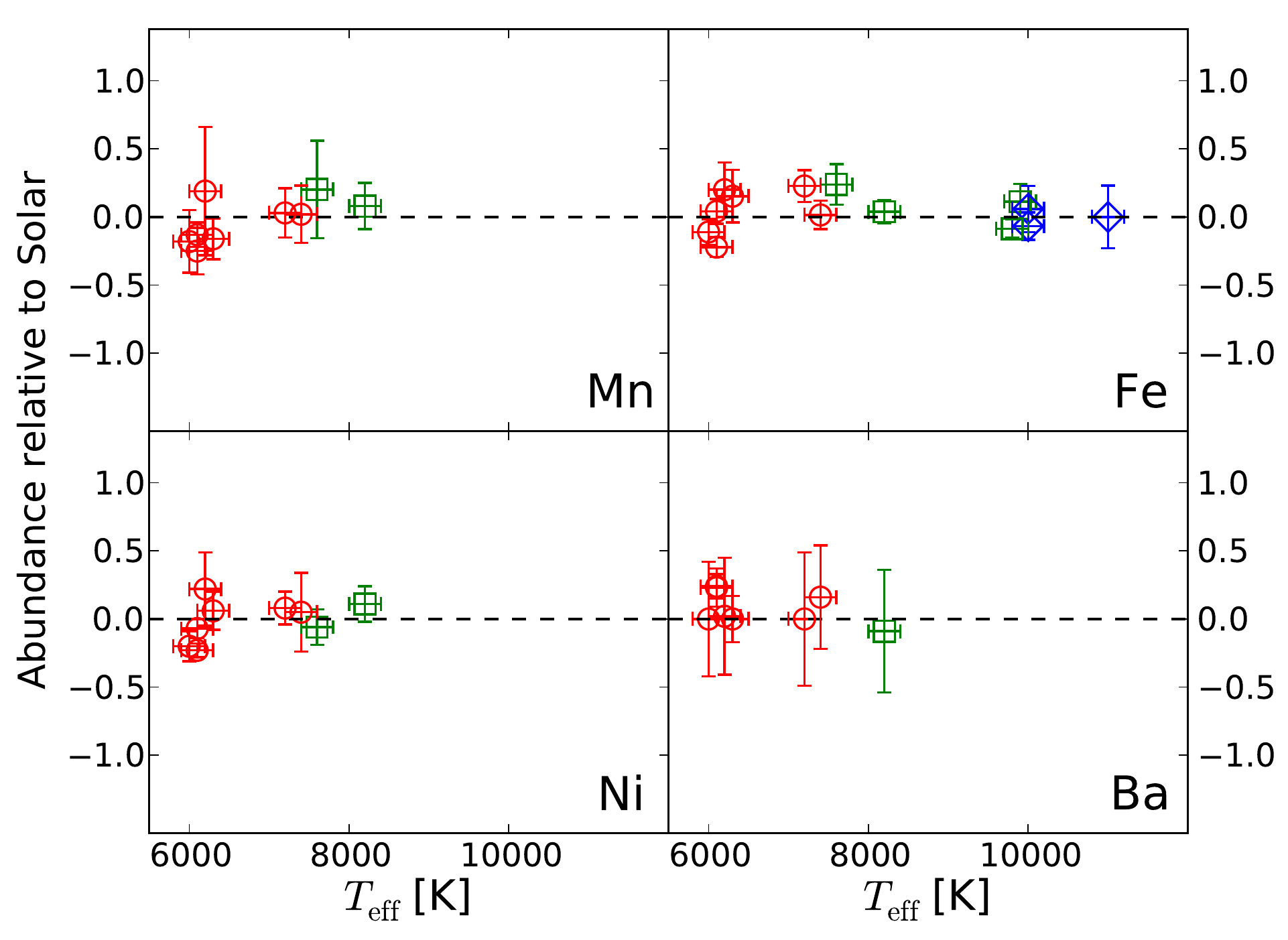}
     \caption{ Same as Fig. \ref{fig:Teff1}, but for Mn, Fe, Ni and 
     Ba.}
     \label{fig:Teff3}
\end{figure}
\begin{figure}
\centering
   \includegraphics[width=\columnwidth]{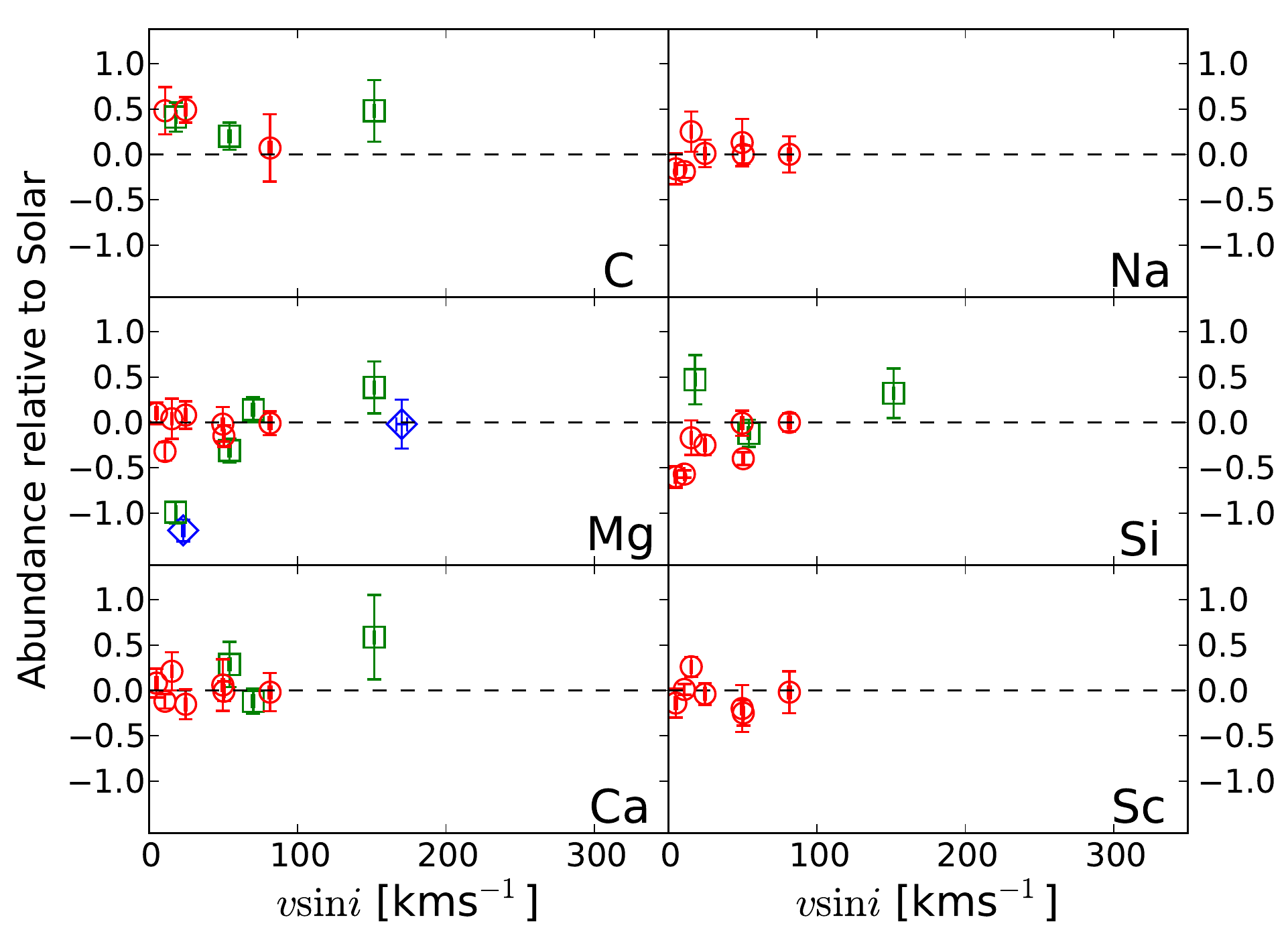}
     \caption{The abundances of C, Na, Mg, Si, Ca and Sc relative to the solar
     abundance \citep{Asplund09} against $v \sin i$. For F- (red circles), 
     A-(green squares) and B- (blue diamonds) type stars.}
     \label{fig:vsini1}
\end{figure}
\begin{figure}
\centering
   \includegraphics[width=\columnwidth]{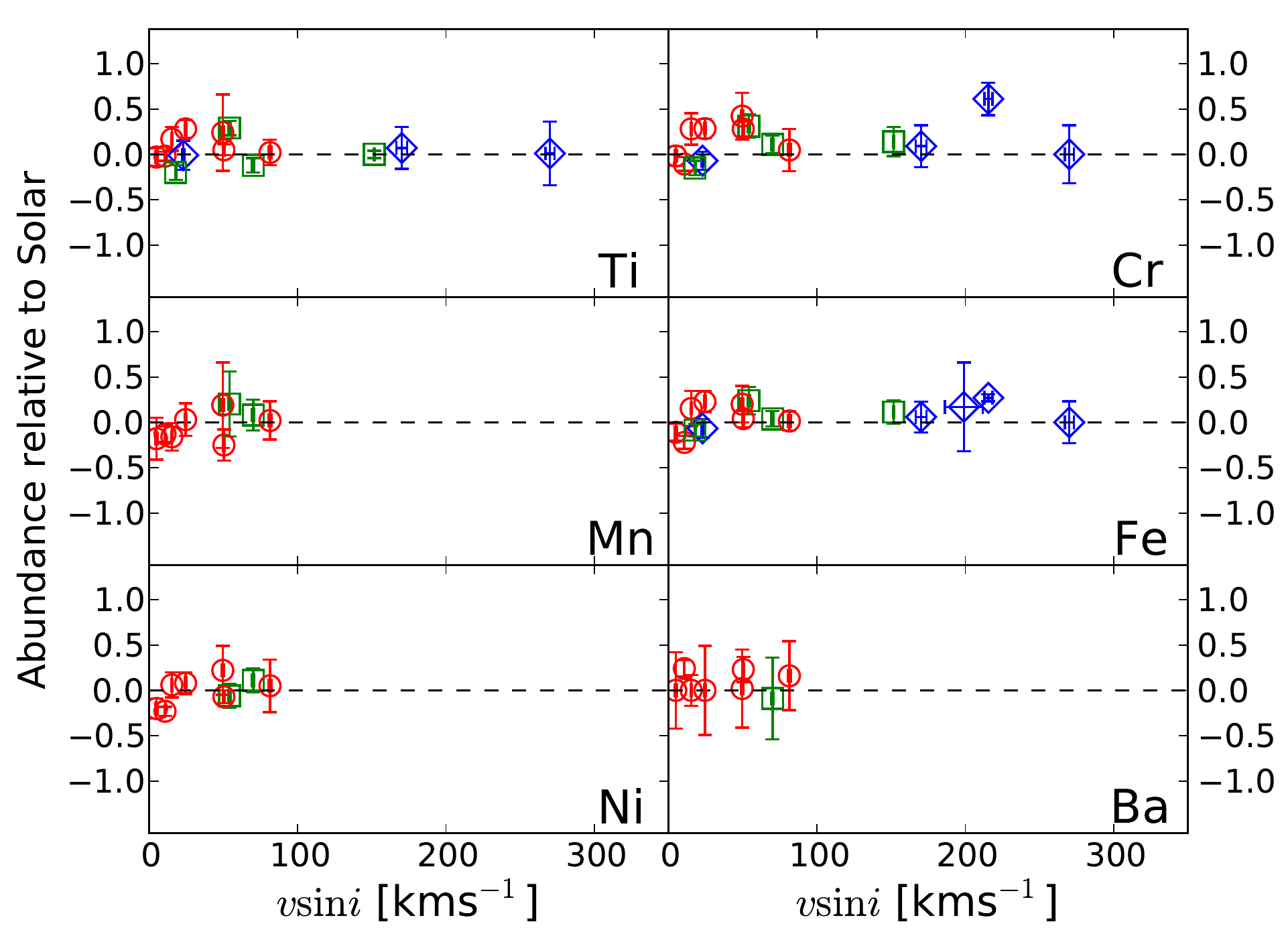}
     \caption{ Same as Fig.~\ref{fig:vsini1}, but for Ti, Cr, Mn, Fe, Ni and 
     Ba.}
     \label{fig:vsini2}
\end{figure}

\subsection{Analysis of chemical abundances}

Fig. \ref{fig:solar} shows the mean abundance of each element obtained 
for the F-, A- and B-type stars. The
error bars are calculated as the standard deviation about the
mean abundance.  We consider only the
measurements from Table~\ref{table:results} with maximum errors 
smaller than 0.5.

To determine whether there is any correlation with the stellar fundamental 
parameters, we have compared each set of element abundances with $T_{\rm eff}$, 
$v \sin i$, $M/$M$_\odot$ and fractional main-sequence age. 
In Figs \ref{fig:Teff1}--\ref{fig:Teff3}, we show abundance as a 
function  $T_{\rm eff}$ and in Figs \ref{fig:vsini1} and \ref{fig:vsini2}, we show abundance 
as a function of $v \sin i$. After comparison between abundance and 
each of the fundamental parameters, we see no statistically 
significant patterns. This is consistent with the findings of \citet{Kilicoglu2016}.

In addition, we have compared our results with the previous 
studies of the open clusters NGC6405, NGC\,5460 and Praesape performed by 
\citet{Kilicoglu2016}, \citet{Fossati11a} and \citet{Fossati07,Fossati08b,Fossati10}, respectively. 
 This allows us to determine whether there is any evidence for correlation between 
 cluster age and abundance. We compare our results with only 
 these clusters since they have all been analysed 
 within this project, and the 
analysis has been either fully carried out (Praesape and NGC\,5460) 
or supervised by one of us (LF)  
(NGC\,6405 and NGC\,6250), to minimize the possibility of 
systematic differences between the results. To compare the results 
from each cluster analysis, we have offset the abundance values of the individual chemical elements according to the cluster metallicities as estimated from Fe abundances of the cluster F and later type stars, which should be less affected by diffusion than earlier type stars.  \\
 
\begin{figure}
\centering
   \includegraphics[width=\columnwidth]{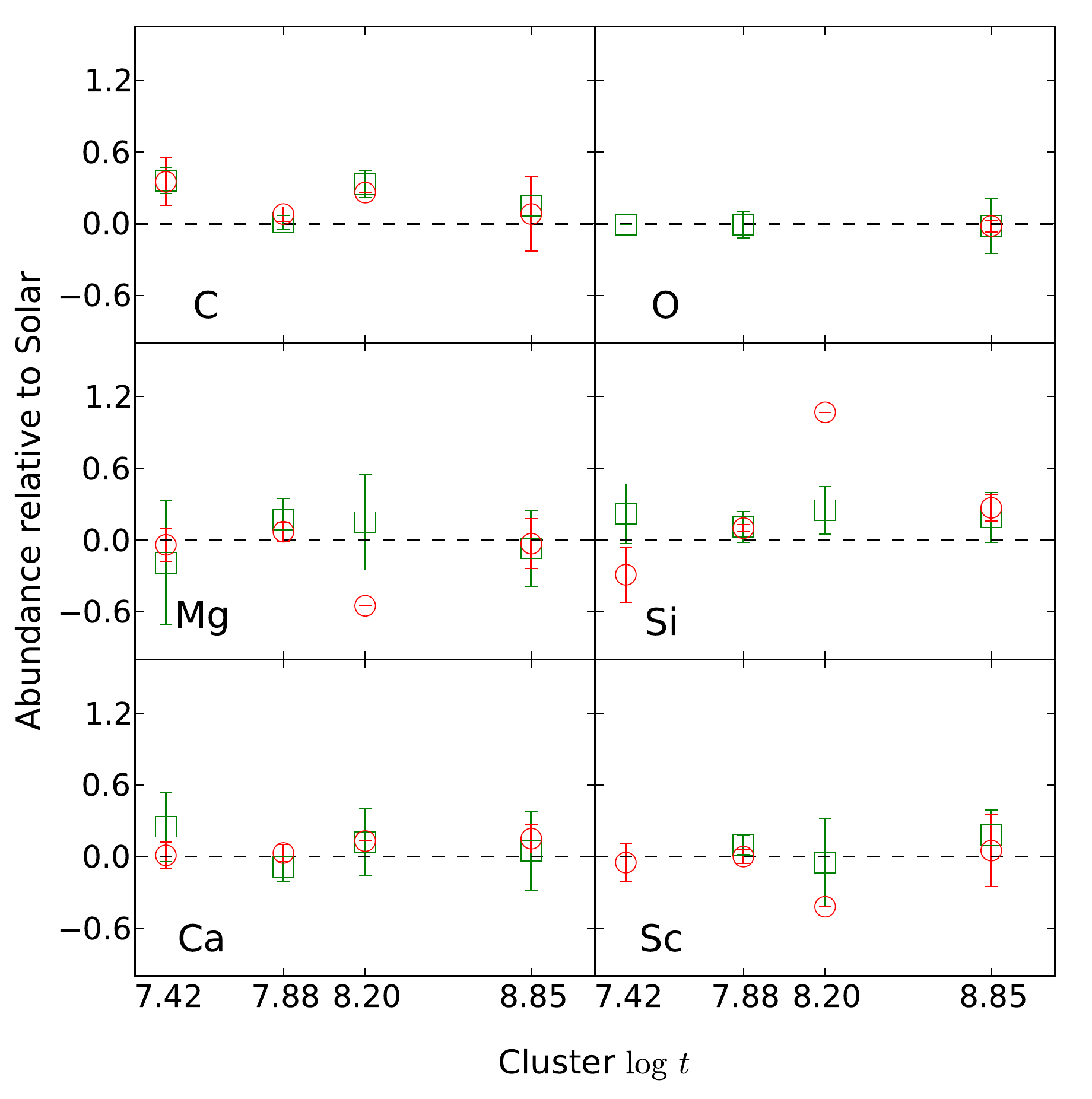}
     \caption{ A comparison between the mean C, O, Mg, Si, Ca and Sc
      abundances found 
     for each of the previous studies and those found in this paper. 
     Mean abundances for 
     F- (red circles) and A-type (green squares) stars are 
     plotted against cluster age ($\log t=7.42$ for NGC\,6250; $\log t=7.88$ for NGC\,6405; 
     $\log t=8.20$ for NGC5460; and $\log t=8.85$ for Praesepe). The error is given by the standard 
     deviation of all the measured abundances.}
     \label{fig:All_Elems_1}
\end{figure}
\begin{figure}
\centering
   \includegraphics[width=\columnwidth]{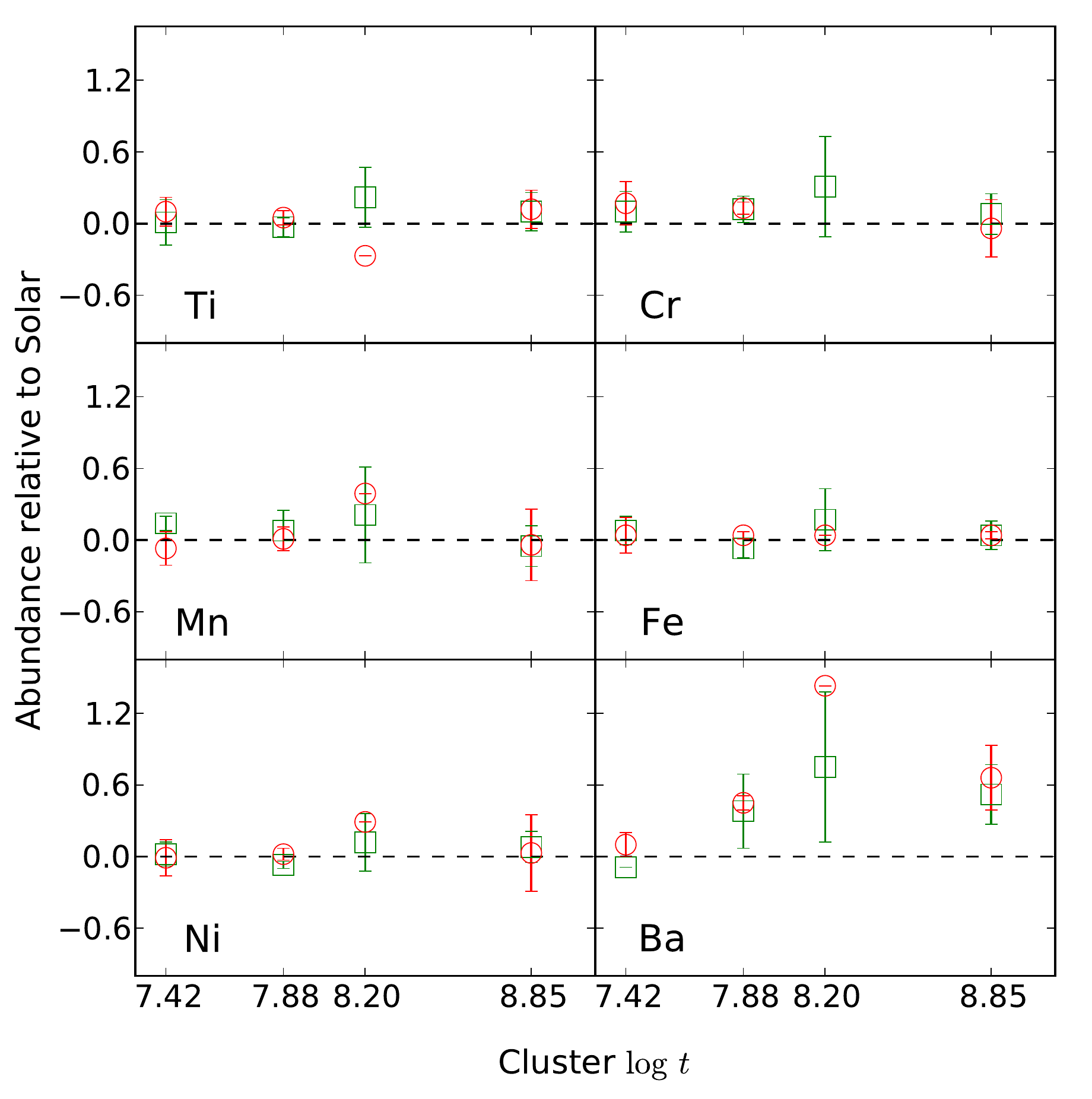}
     \caption{ Same as Fig. \ref{fig:All_Elems_2} but for Ti, Cr, 
     Mn, Fe, Ni and Ba.}
     \label{fig:All_Elems_2}
\end{figure}

In NGC\,6250, we found that O, Na, Sc, Ti, Cr, Mn, Ni, Zn and Y all have solar 
abundances within the uncertainties, while S and V are 
overabundant. These results are consistent with the findings of 
\citet{Fossati08b}, \citet{Fossati11a} and \citet{Kilicoglu2016} for 
the Praesepe cluster, NGC\,5460 and NGC\,6405, respectively (see
Figs \ref{fig:All_Elems_1} and \ref{fig:All_Elems_2}). \\

Similarly to what was found by \citet{Fossati08b}, \citet{Fossati11a} and \citet{Kilicoglu2016}
 in the Praesepe cluster, NGC\,5460 and NGC\,6405, we have found an overabundance of C in the F- and A-type stars of NGC 6250. However, we do not see any trend with age (see
Fig. \ref{fig:All_Elems_1}).

For all of the F-type stars, we find a solar abundance of Mg, for the A-type 
and B-type stars there is an underabundance of Mg; however, there is a 
large spread in the results and all but two stars have approximately solar 
abundance, which matches with the results of the previous studies.

In agreement with previous studies, we found that in A-type stars Si is overabundant; 
however, at odds with previous studies, we found that in F-type stars Si is underabundant. 
Fig. \ref{fig:Teff1} indicates the presence of a possible correlation between $T_{\rm eff}$ and the Si abundance, though a further analysis reveals that this apparent correlation is not statistically significant.

For all of the F-type stars, we find a solar abundance of Ca that is 
consistent with the previous results. However, for the A-type 
stars, we find an overabundance, which is contrary to the findings of the previous 
studies; the origin of this is unclear. 

We measure the abundance of Fe in all of our stars to be 
approximately solar. 
For both Mn and Fe, \citet{Fossati11a} found an increase in 
 abundance with $T_{\rm eff}$,  which we do not; 
this therefore may be the result of an age effect. 
The narrow $T_{\rm eff}$ range of the 
stars analysed by  \citet{Fossati08b} for the Praesepe cluster means we are 
unable to provide any definite conclusions until the remaining clusters 
are analysed.

We measure an almost solar abundance for Ba,  albeit with relatively large uncertainties. 
This is in contrast with the findings of \citet{Fossati08b}, \citet{Fossati11a} 
and \citet{Kilicoglu2016} who all report overabundances. To 
understand each of the results together, we plot the mean abundance of Ba
measured for each cluster in Fig. \ref{fig:All_Elems_2}. We did not consider the 
stars HD\,122983 and 
HD\,123182 from NGC\,5460 because of their apparent chemical peculiarities \citep{Fossati11a}. 
From Fig. \ref{fig:All_Elems_2}, we obtain a hint of a positive 
correlation of Ba abundance with age; however,  
the abundance uncertainties are too large to draw any concrete conclusion. By 
analysing further clusters we will be able to determine whether this effect is the result of 
diffusion or the different chemistry of the star-forming region for each 
cluster.

We measure Nd to be overabundant in four stars; however, the 
data from previous papers are too sparse to provide any conclusion.\\

Finally, we have compared our results with the study of chemically peculiar 
magnetic Ap stars by \citet{Bailey2014}. This allows us to examine the 
differences and similarities between abundance trends of chemically 
normal and chemically peculiar stars. \citet{Bailey2014} found 
statistically significant trends between He, Ti, Cr, Fe, Pr and Nd 
and stellar age. They also found a strong trend between the abundances of 
Cr and Fe, and $T_{\rm eff}$. For Cr, an underabundance was observed for stars 
with $T_{\rm eff} \lesssim 7000\,$K, for stars with 
$T_{\rm eff} \gtrsim 7000\,$K the abundance of Cr 
sharply rises and peaks at $T_{\rm eff} \sim 10000\,$K before falling back to 
approximately solar. For Fe, an underabundance was observed for stars 
with $T_{\rm eff} \lesssim 8000\,$K and an overabundance for the 
remaining stars. These results are in stark contrast with what we observed for 
NGC\,6250. This suggests that the abundance of 
chemical elements in the photosphere of chemically 
normal F-, A- and B-type stars remains relatively constant during their main-sequence 
lifetime except when influenced by a magnetic field.

\section{Conclusions}\label{Sect_Conclusions}
We have presented the new code for spectral analysis,
\spartisimplex. Based on \cossamsimplex, a modified version of the
radiative transfer code \cossam, \spartisimplex\ employs the inversion
algorithm LMA and allows one to recover the
abundance of the chemical elements of non-magnetic stellar atmospheres. To test
our new code, we have performed the abundance analysis of
the Sun, HD\,32115 and 21\,Peg and compared our results with those 
previously published in the thorough works by  \citet{Prsa2016}, 
\citet{Asplund09}, \citet{Fossati2011b} and \citet{Fossati09}, finding 
 excellent agreement.

We have applied our new code for a spectroscopic study of the open
cluster NGC\,6250, which was observed with the FLAMES instrument of
the ESO VLT. From the observed sample of stars, we have performed
cluster membership analysis based on a $K$-means clustering procedure 
and analysis of the photometry.
As a result of our analysis, we selected  19 stars from our sample as
members of the cluster.  We have computed the cluster mean proper
motions of 0.4\,$\pm$\,3.0\,mas\,yr$^{-1}$ in RA,
$-$4.80\,$\pm$\,3.2\,mas\,yr$^{-1}$ in DEC and radial velocity of
$-$10\,$\pm$\,6\,km\,s$^{-1}$. These values agree within the 
errors with the values calculated by \citet{Kharchenko13}. The age and 
distance given by \citet{Kharchenko13} agree well with our 
photometric analysis of the cluster.
 
 Finally, we have examined the chemical abundance measurements 
for each star and searched for any trend between abundance and 
the stellar fundamental parameters and between the abundance 
measured in this study and the abundance measured in the 
previous studies of older clusters by \citet{Fossati07,Fossati08b,Fossati10}, 
\citet{Fossati11a} and \citet{Kilicoglu2016}. 
Our results for the abundance of 
O, Na, Sc, Ti, Cr, Mn, Ni, Zn and Y are solar 
 within the uncertainties, while S and V are 
overabundant. These results are consistent with 
previous studies. We do not find evidence of the correlation 
between either the Fe or Mn  abundance 
and $T_{\rm eff}$ found 
by \citet{Fossati11a}; however, this may be evidence of an age 
effect and we need to study more clusters before being able to 
determine this.
We find hints of an increase in mean Ba abundance with cluster 
age but more clusters should be analysed to confirm this trend.
Comparing our results with those from \citet{Bailey2014}, who 
searched for trends between chemical abundances and 
stellar parameters of chemically peculiar magnetic Ap stars, 
suggests that the abundance of 
chemical elements in the photosphere of chemically 
normal F-, A- and B- 
type stars remains relatively constant during their main-sequence 
lifetime except when influenced by a magnetic field.
\section*{Acknowledgements}
This paper is based on observations made with ESO Telescopes at 
the Paranal Observatory under programme ID 079.D-0178.
We thank Claudia Paladini for the re-reduction of the UVES spectra.
AM acknowledges the support of a Science and Technology Facilities 
Council (STFC) PhD studentship. Thanks go to AdaCore for providing 
the GNAT GPL Edition of its Ada compiler. This publication makes use of data products from the AAVSO Photometric All Sky Survey (APASS). Funded by the Robert Martin Ayers Sciences Fund and the National Science Foundation. We thank the referee 
Charles Proffitt for providing constructive comments that led to a significant 
improvement of the manuscript.



\bibliographystyle{mnras}
\bibliography{CaPaper} 

\bsp	
\label{lastpage}
\end{document}